\documentclass[a4paper, amsfonts, amssymb, amsmath, reprint, showkeys, nofootinbib, twoside, apsprl]{revtex4-1}
\usepackage[letterpaper,top=2cm,bottom=2cm,left=2cm,right=2cm,marginparwidth=1.75cm]{geometry}
\pdfoutput=1

\usepackage{amsmath}
\usepackage{amsfonts}
\usepackage{amssymb}

\usepackage{enumitem}

\usepackage{tikz}

\usetikzlibrary{spy}
\usetikzlibrary{decorations.pathreplacing}
\usetikzlibrary{decorations.pathmorphing}
\usetikzlibrary{circuits.logic.US,circuits.logic.IEC,fit}

\usepackage{array}
\newcommand\addvmargin[1]{ \node[fit=(current bounding box),inner ysep=#1,inner xsep=0]{}; }
\usepackage{todonotes}
\usepackage[normalem]{ulem}
\usepackage{todonotes}
\usepackage{hyperref}
\usepackage{cleveref}

\definecolor{good_red}{RGB}{212, 17, 89}
\definecolor{good_blue}{RGB}{26, 133, 255}

\newcommand{\tikzscale}{0.5}
\newcommand{\latticecolor}{gray}
\newcommand{\latticethickness}{thick}
\newcommand{\supportthickness}{2pt}

\newcommand{\op}[1]{\operatorname{#1}}

\begin{document}
\title{Tile Codes: High-Efficiency Quantum Codes on a Lattice with Boundary}

\author{Vincent Steffan$^\dagger$\textsuperscript{1}
}
\email[Correspondence email address: ]{vincent.steffan@meetiqm.com}
\author{Shin Ho Choe$^\dagger$\textsuperscript{1}}

\author{Nikolas P.\ Breuckmann\textsuperscript{2}}
\author{Francisco Revson Fernandes Pereira\textsuperscript{1}}

\author{Jens Niklas Eberhardt\textsuperscript{3}}

\affiliation{\textsuperscript{1}IQM Quantum Computers, Georg-Brauchle-Ring 23-25, 80992 Munich, Germany}
\affiliation{\textsuperscript{2}Breuqmann Ltd., Redcross Village, BS2 0BB Bristol, United Kingdom}
\affiliation{\textsuperscript{3}Institute of Mathematics, Johannes Gutenberg-Universität Mainz, Germany}

\date{\today} 

\def\thefootnote{$\dagger$}\footnotetext{These authors contributed equally to this work}\def\thefootnote{\arabic{footnote}}

\begin{abstract}
We introduce \textit{tile codes}, a simple yet powerful way of constructing quantum codes that are local on a planar 2D-lattice. Tile codes generalize the usual surface code by allowing for a bit more flexibility in terms of locality and stabilizer weight. Our construction does not compromise on the fact that the codes are local on a lattice with open boundary conditions. Despite its simplicity, we use our construction to find codes with parameters $[[288, 8, 12]]$ using weight-6 stabilizers and $[[288, 8, 14]]$ using weight-8 stabilizers, outperforming all previously known constructions in this direction. Allowing for a slightly higher non-locality, we find a $[[512, 18, 19]]$ code using weight-8 stabilizers, which outperforms the rotated surface code by a factor of more than 12. Our approach provides a unified framework for understanding the structure of codes that are local on a 2D planar lattice and offers a systematic way to explore the space of possible code parameters. In particular, due to its simplicity, the construction naturally accommodates various types of boundary conditions and stabilizer configurations, making it a versatile tool for quantum error correction code design.
\end{abstract}

\maketitle

The surface code was introduced in 1998~\cite{bravyi1998quantum,freedman2001projective} and until today, it is arguably the quantum error-correcting code best suited for current and near-term quantum devices~\cite{Andersen_2020,Krinner_2022,google2023, google2024}. 
It owes its practicability mainly to the fact that its stabilizer checks are local on a regular, planar 2D lattice. In fact, having nearest-neighbor connectivity between the physical qubits on a chip is sufficient to realize a surface code without further overhead. Fundamentally, this locality comes at a price: The surface code encodes just one logical qubit. 

Compared to the surface code, so-called quantum Low-Density Parity-Check (qLDPC) codes may offer higher encoding rates and better protection of encoded qubits~\cite{breuckmannQuantumLowDensityParityCheck2021a}.
However, realizing the promises of qLDPC codes is far from trivial: qLDPC codes with the best theoretical performance are inherently non-2D~\cite{Bravyi_2010}.

Recent advances in quantum hardware design, particularly for superconducting circuits, suggest that slightly relaxed locality constraints may be achievable in near-term devices \cite{vigneau2025quantumerrordetectionqubitresonator, renger2025superconductingqubitresonatorquantumprocessor, Wallraff2004, blais2004, Song2017, Song2019}.
This raises hopes, but it is a priori not clear how to leverage this additional flexibility to develop more efficient codes. 

This Letter addresses the main question of constructing more efficient QEC codes by slightly relaxing the locality of the surface code. We introduce \textit{tile codes}, a natural generalization of the surface code that relaxes locality in two key ways. Instead of assuming nearest-neighbor connectivity, we allow for the stabilizer generators are supported in boxes of a certain size. Second, we allow for stabilizer weights higher than 4, as in the case of the surface code. Tile codes follow a systematic approach where stabilizer tiles (specific stabilizer patterns) are placed on a layout according to simple rules that guarantee that the stabilizers of the code commute.

While our general construction is extremely simple, we discover codes that outperform the surface code by factors of up to 12 while still maintaining practical locality constraints. For example, with just a 3×3 box size and weight-6 stabilizers, we obtain a ${[[288,8,12]]}$ code---a 4-times improvement in efficiency compared to the rotated surface code. Increasing to weight-8 stabilizers yields a ${[[288,8,14]]}$ code, while a 4×4 box with weight-8 stabilizers produces a ${[[288,18,13]]}$ and ${[[512, 18, 19]]}$ code---a remarkable 10-times and 12-times gain in efficiency compared to the rotated surface code. 
We remark that the ${[[288,8,12]]}$ code was first discovered by Liang~et~al.~\cite{yuannewarticle} and served as an inspiration for our constructions.

Tile codes are related to Bivariate Bicycle (BB) codes~\cite{linQuantumTwoblockGroup2023,bravyi2023highthreshold,eberhardt2024pruningqldpccodesbivariate,eberhardt2024logicaloperatorsfoldtransversalgates,chen2025anyontheorytopologicalfrustration, liang2025generalizedtoriccodestwisted, liang2024operatoralgebraalgorithmicconstruction}. BB codes generalize the toric code, allowing for some modest (constant) amount of non-locality assuming periodic boundaries on the lattice.
This means the stabilizer checks are not truly local in a planar layout.
There is still a growing number of stabilizer checks that require a growing amount of non-locality.
This issue is addressed if one can build reliable long-range couplers~\cite{bravyi2023highthreshold} or by simulating non-local interactions by teleportation-based routing~\cite{Berthusen_2025} or swapping qubits~\cite{pattison2023hierarchicalmemoriessimulatingquantum}. 
Tile codes, on the other hand, do not require periodicity and thus have true $O(1)$-locality on a planar 2D lattice. 
With that, tile codes address this major challenge in quantum error correction, achieving high efficiency while preserving implementability on near-term hardware.

\begin{figure*}
    \begin{tikzpicture}[spy using outlines = {circle, magnification = 1.65, size = 3.5cm}, scale = .7]
        \begin{scope}[scale=0.6]

        \foreach \x in {0,...,11} {
            \draw[\latticecolor, \latticethickness] (\x,0) -- (\x,12);
        }
        
        \foreach \y in {0,...,11} {
            \draw[\latticecolor, \latticethickness] (0,\y) -- (12,\y);
        }

        \begin{scope}[xshift = 4cm, yshift = 1cm]
        \def\horizontalqubits{(0,0),(2,1), (2,2)}
        \def\verticalqubits{(0,2),(1,2),(2,0)}

        \draw[color = good_red,thick] (0,0) circle (0.3);
        \draw[color = good_red, line width = \supportthickness] (0,0) -- (1,0);
        \draw[color = good_red, line width = \supportthickness] (2,1) -- (3,1);
        \draw[color = good_red, line width = \supportthickness] (2,2) -- (3,2);
        \draw[color = good_red, line width = \supportthickness] (0,2) -- (0,3);
        \draw[color = good_red, line width = \supportthickness] (1,2) -- (1,3);
        \draw[color = good_red, line width = \supportthickness] (2,0) -- (2,1);

        \fill[opacity = .1, color = good_red, line width = \supportthickness] (0,0) -- (3,0) -- (3,3) -- (0,3) -- cycle;
        \end{scope}

        \def\horizontalqubits{(-1,5),(-1,6), (1,6)}
        \def\verticalqubits{(-1,7),(0,7),(1,5)}

        \draw[color = good_blue,thick] (-1,5) circle (0.3);
        \draw[color = good_blue, line width = \supportthickness,densely dotted] (-1,7) -- (0,7);
        \draw[color = good_blue, line width = \supportthickness] (0,5) -- (1,5);
        \draw[color = good_blue, line width = \supportthickness] (1,5) -- (2,5);
        \draw[color = good_blue, line width = \supportthickness, densely dotted] (-1,5) -- (-1,6);
        \draw[color = good_blue, line width = \supportthickness,densely dotted] (-1,6) -- (-1,7);
        \draw[color = good_blue, line width = \supportthickness] (1,7) -- (1,8);

        \fill[opacity = .1, color = good_blue] (-1,5) -- (2,5) -- (2,8) -- (-1,8) -- cycle;

        \foreach \x in {0,...,9} {
            \foreach \y in {0,...,9} {
                \draw[fill=black] (\x,\y) circle (0.15);
            }
        }
        
        \foreach \y in {0,...,9} {
            \foreach \x in {-2,-1} {
                \draw[fill=good_blue, color = good_blue] (\x,\y) circle (0.15);
            }
        }
        
        \foreach \y in {0,...,9} {
            \foreach \x in {10,11} {
                \draw[color = good_blue, fill=good_blue] (\x,\y) circle (0.15);
            }
        }

        \foreach \x in {0,...,9} {
            \foreach \y in {10,11} {
                \draw[color = good_red, fill=good_red] (\x,\y) circle (0.15);
            }
        }

        \foreach \x in {0,...,9} {
            \foreach \y in {-2,-1} {
                \draw[color = good_red, fill=good_red] (\x,\y) circle (0.15);
            }
        }
        
        \fill[opacity = .1, color = good_red] (9,10) -- (9,13) --  (12,13) -- (12,10) -- cycle;
        \fill[opacity = .1, color = good_blue] (10,9) -- (13,9) --  (13,12) -- (10,12) -- cycle;


        \spy[black, connect spies, very thick] on (2.3,1.05) in node at (-4.2,1.2);
        \node[scale = 1.2] at (-10,-2) {\textbf{(a)}};

        \spy[black, connect spies] on (0.2,2.75) in node at (-8.5,4.5);
        \node[scale = 1.2] at (-18,4) {\textbf{(b)}};
        
        \spy[black, connect spies] on (4.45,4.45) in node at (10.5,3);
        \node[scale = 1.2] at (15,0.7) {\textbf{(c)}};
        \end{scope}
    \end{tikzpicture}
    \caption{An example of a $[[288,8,12]]$ tile code following our construction. The qubits sit on the edges of the lattice. The $\op X$- and $\op Z$-type stabilizer tiles sit at the vertices of the lattice. A black dot indicates that both an $\op X$- and a $\op Z$-type stabilizer tile is there. Red dots and blue dots indicate a $\op X$-type and $\op Z$-type stabilizer, respectively. The dots are the anchors of the $3\times 3$ boxes that contain the stabilizer tiles. In \textbf{(a)}, we zoom in on an example of an $\op X$-type bulk stabilizer tile, and in \textbf{(b)} an example of a $\op Z$-type boundary stabilizer tile. Since the dotted edges that usually would be in the support of the $\op Z$-type stabilizer are not in the set of physical qubits, the support of this stabilizer is therefore restricted to the non-dotted edges for this tile code. In \textbf{(c)}, we visualize that naturally, if two boundary stabilizers overlap, by the choice of the layout, their overlap is contained in the lattice of physical qubits. We intentionally only draw the $B \times B$ box containing the stabilizers and do not mark the edges in the support of the stabilizers since the argument is agnostic of the specific stabilizer.}\label{fig: masterfigure}
\end{figure*}
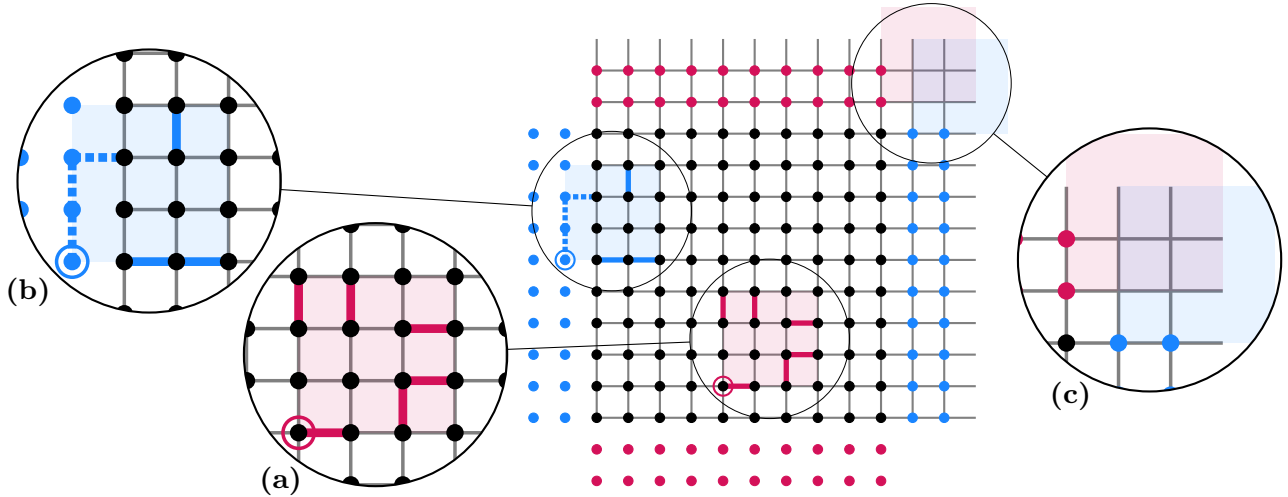

\textit{Tile codes ---}
We now introduce a family of quantum stabilizer codes which we call \emph{tile codes}. For a visual overview, we refer to~\Cref{fig: masterfigure}.
Tile codes are a natural generalization of surface codes:
similar to surface codes, they have a bulk region with translation invariant stabilizer checks and boundaries where stabilizer checks can have restricted supports.
They arise by choosing a \textit{local structure} which consists of shapes of a $\op X$- and $\op Z$-type stabilizer called tiles, and a \textit{global structure} which is a layout of qubits and positions of stabilizer checks.

A \emph{tile} is a subset of edges of a $B\times B$-square grid, excluding the edges on the top row and the rightmost column.
We identify edges with physical qubits so that a tile defines the support of a Pauli-$\op X$ or Pauli-$\op Z$ stabilizer check.
For example, one can recover the vertex and plaquette operators of the surface code, which are supported on the edges adjacent to a vertex and a face, by using the following tiles $\op{X}$- and $\op{Z}$-tile: 
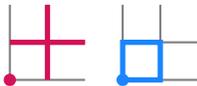
\begin{equation*}\label{eq: surfacecodestabs}
    \begin{tikzpicture}[scale=\tikzscale, baseline=(current bounding box.center)]
    \begin{scope}[xshift=3cm]
        \draw[\latticecolor, \latticethickness] (0,2) -- (0,0) -- (2,0) ;
        \draw[\latticecolor, \latticethickness] (0,1) -- (2,1);
        \draw[\latticecolor, \latticethickness] (1,0) -- (1,2);
        \draw[good_blue, line width = \supportthickness] (0,0) -- (0,1) -- (1,1) -- (1,0) -- (0,0);

        \draw[color = good_blue, fill = good_blue] (0,0) circle (0.15);
        \end{scope}
        \begin{scope}[xshift=0cm]
            \draw[\latticecolor, \latticethickness] (0,2) -- (0,0) -- (2,0) ;
        \draw[\latticecolor, \latticethickness] (0,1) -- (2,1);
        \draw[\latticecolor, \latticethickness] (1,0) -- (1,2);
        \draw[good_red,  line width = \supportthickness] (0,1) -- (2,1);
        \draw[good_red,  line width = \supportthickness] (1,0) -- (1,2);

        \draw[color = good_red, fill = good_red] (0,0) circle (0.15);
        \end{scope}
    \end{tikzpicture}
\end{equation*}

We will draw tiles with a marked dot which indicates their placement in the layout that we will explain next.

The \emph{layout} of a tile code determines the full set of physical qubits, as well as the position of tiles (and hence stabilizer checks).
More specifically, a layout is a finite subset of the square grid.
The subset of edges is identified with the physical qubits.
The vertices are marked depending on the placement of tiles which represent the stabilizer checks.
We will mark vertices in black, red, or blue.
We place both $\op X$- and $\op Z$-tiles on black vertices, whereas, on each red vertex of the layout, we place only an $\op X$-tile and on each blue vertex only a $\op Z$-tile. 
Note that not all vertices have to be marked.
Unmarked vertices do not have any stabilizer checks associated with them.
If the set of edges of the tile code does not contain the whole support of a stabilizer, then the support of the stabilizer should be thought of as truncated to the available edges/qubits. We will refer to stabilizers whose support is contained in the set of edges of the layout as \textit{bulk stabilizers} and to the remaining stabilizers as \textit{boundary stabilizers}.

This concept of a layout is hence a generalization of the various shapes and sizes of surface codes, both rotated and unrotated,  and the notion of smooth and rough boundaries. For example, using the surface code tiles, the following layout specifies a usual, unrotated surface code with distance 4:

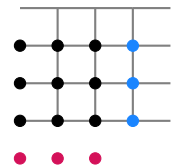
\begin{equation*}\label{eq: surfacecodelattice}
  \begin{tikzpicture}[scale=\tikzscale, baseline=(current bounding box.center)]
      \begin{scope}[xshift=0cm]
      \foreach \y in {0,1,2,3} {
          \draw[\latticecolor, \latticethickness] (0,\y) -- (3,\y);
          \draw[\latticecolor, \latticethickness] (3,\y) -- (4,\y);
      }
      
      \foreach \x in {1,2,3} {
          \draw[\latticecolor, \latticethickness] (\x,0) -- (\x,3);
      }
      
      \foreach \x in {0,1,2} {
          \foreach \y in {0,1,2} {
              \draw[fill=black] (\x,\y) circle (0.15);
          }
      }
      
      \foreach \y in {0,1,2} {
          \draw[good_blue, fill=good_blue] (3,\y) circle (0.15);
      }
      
      \foreach \x in {0,1,2} {
          \draw[good_red, fill=good_red] (\x,-1) circle (0.15);
      }
      \end{scope}
  \end{tikzpicture}
\end{equation*}
While this is an unorthodox way of presenting the surface code, our convention will be useful for the general construction of tile codes which we describe in the following.

\textit{Systematic construction of tile codes ---} 
Not every choice of tiles and layout yields a well-defined CSS code. Indeed, one of the key challenges in constructing tile codes is to choose these such that the boundary stabilizers commute. We will now explain a systematic construction that ensures commutativity.

We first determine the local structure of our code by choosing an $\op{X}$-tile and a $\op{Z}$-tile such that

\begin{enumerate}[label={(T\arabic*)}]
    \item both tiles are confined within a box of size $B\times B$ with no support at the top horizontal and right vertical edges of the box, and,
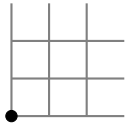
\label{it:condition for size tiles bb code}
    \begin{equation*}
    \begin{tikzpicture}[scale=\tikzscale]

            \draw[\latticecolor, \latticethickness] (0,0) -- (3,0);
            \draw[\latticecolor, \latticethickness] (0,0) -- (0,3);
            \draw[\latticecolor, \latticethickness] (1,0) -- (1,3);
            \draw[\latticecolor, \latticethickness] (2,0) -- (2,3);
            \draw[\latticecolor, \latticethickness] (0,1) -- (3,1);
            \draw[\latticecolor, \latticethickness] (0,2) -- (3,2);
            
            \draw[fill=black] (0,0) circle (0.15);

    \end{tikzpicture}
\end{equation*}
    \item the $\op X$-tile and $\op Z$-tile determine each other in the following  way: for each $\op X$-tile with support on a horizontal (resp.\ vertical) edge with coordinate $(x,y)$, there is a vertical (resp.\ horizontal) edge with coordinates $(B-1-x, B-1-y)$ in the $\op Z$-tile. \label{it:condition for tiles bb code}
\end{enumerate}
The condition \ref{it:condition for tiles bb code} guarantees that $\op X$- and $\op Z$-tiles, in any relative position, have an even overlap.\footnote{On the infinite lattice $\mathbb{Z} \times \mathbb{Z}$, tiles with condition (T2) give rise to a Pauli Hamiltonian with commuting terms. It turns out~\cite{Haah_2017, chen2025anyontheorytopologicalfrustration} that the groundspace of this Hamiltonian exhibits topological order only if (T2) holds. Practically this means that (T2) is neccessary if we want that our tile codes have distance scaling with the dimensions of the layout.}

Next, we start constructing a layout by defining a designated set of bulk-stabilizer positions. While this set can be arbitrary, we will always restrict ourselves to following (rotated) rectangular shapes.

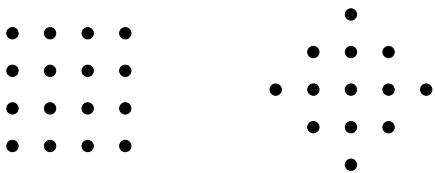
\begin{equation*}\label{eq: bulkstablattices}
  \begin{tikzpicture}[scale = \tikzscale, baseline=(current bounding box.center)]
          \foreach \x in {0,...,3} {
              \foreach \y in {0,...,3} {
                  \draw[fill=black] (\x,\y) circle (0.15);
              }
          }
                  \begin{scope}[xshift=9cm, yshift = 1.5cm]
          \draw[fill=black] (-2,0) circle (0.15);
          \draw[fill=black] (-1,0) circle (0.15);
          \draw[fill=black] (-1,-1) circle (0.15);
          \draw[fill=black] (-1,1) circle (0.15);
          \draw[fill=black] (0,-2) circle (0.15);
          \draw[fill=black] (0,-1) circle (0.15);
          \draw[fill=black] (0,0) circle (0.15);
          \draw[fill=black] (0,1) circle (0.15);
          \draw[fill=black] (0,2) circle (0.15);
          \draw[fill=black] (1,0) circle (0.15);
          \draw[fill=black] (1,-1) circle (0.15);
          \draw[fill=black] (1,1) circle (0.15);
          \draw[fill=black] (2,0) circle (0.15);
      \end{scope}
  \end{tikzpicture}
\end{equation*}

Since in the end, we want these stabilizers to be bulk stabilizers and therefore their support completely contained in the layout, we will choose as the set of edges the union of all $B\times B$ boxes over all bulk stabilizer positions. For example, the following layouts are induced by the lattices depicted above together with $3\times 3$ boxes containing the bulk stabilizers.
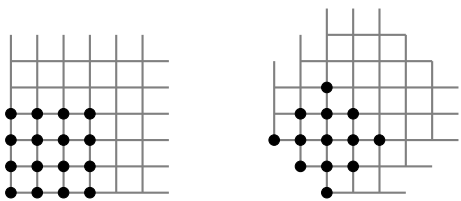
\begin{equation*}\label{eq: bulkstablatticesandqubits}
    \begin{tikzpicture}[scale = .7, baseline=(current bounding box.center)]
            \foreach \x in {0,...,5} {
                \draw[\latticecolor, \latticethickness] (\x/2,0) -- (\x/2,3);
            }
            
            \foreach \y in {0,...,5} {
                \draw[\latticecolor, \latticethickness] (0,\y/2) -- (3,\y/2);
            }
            
            3x3 grid of black dots
            \foreach \x in {0,...,3} {
                \foreach \y in {0,...,3} {
                    \draw[fill=black] (\x/2,\y/2) circle (0.1);
                }
            }
                    \begin{scope}[xshift=6cm, yshift = 1cm]
            Vertical lines
            \draw[\latticecolor, \latticethickness] (-2/2, 0) -- (-2/2, 3/2);
            \draw[\latticecolor, \latticethickness] (-1/2, -1/2) -- (-1/2, 4/2);
            \draw[\latticecolor, \latticethickness] (0, -1) -- (0, 5/2);
            \draw[\latticecolor, \latticethickness] (1/2, -1) -- (1/2, 5/2);
            \draw[\latticecolor, \latticethickness] (1, -1) -- (1, 5/2);
            \draw[\latticecolor, \latticethickness] (3/2, -1/2) -- (3/2, 4/2);
            \draw[\latticecolor, \latticethickness] (2, 0) -- (2, 3/2);
            
            \draw[\latticecolor, \latticethickness] (0, -2/2) -- (3/2, -2/2);
            \draw[\latticecolor, \latticethickness] (-1/2, -1/2) -- (4/2, -1/2);
            \draw[\latticecolor, \latticethickness] (-2/2, -0/2) -- (5/2, -0/2);
            \draw[\latticecolor, \latticethickness] (-2/2, 1/2) -- (5/2, 1/2);
            \draw[\latticecolor, \latticethickness] (-2/2, 2/2) -- (5/2, 2/2);
            \draw[\latticecolor, \latticethickness] (-1/2, 3/2) -- (4/2, 3/2);
            \draw[\latticecolor, \latticethickness] (0/2, 4/2) -- (3/2, 4/2);
            
            \draw[fill=black] (-2/2,0/2) circle (0.1);
            \draw[fill=black] (-1/2,0/2) circle (0.1);
            \draw[fill=black] (-1/2,-1/2) circle (0.1);
            \draw[fill=black] (-1/2,1/2) circle (0.1);
            \draw[fill=black] (-0/2,-2/2) circle (0.1);
            \draw[fill=black] (-0/2,-1/2) circle (0.1);
            \draw[fill=black] (-0/2,0/2) circle (0.1);
            \draw[fill=black] (-0/2,1/2) circle (0.1);
            \draw[fill=black] (-0/2,2/2) circle (0.1);
            \draw[fill=black] (1/2,0/2) circle (0.1);
            \draw[fill=black] (1/2,-1/2) circle (0.1);
            \draw[fill=black] (1/2,1/2) circle (0.1);
            \draw[fill=black] (2/2, 0/2) circle (0.1);
        \end{scope}
    \end{tikzpicture}
\end{equation*}

Next, we want to add boundary stabilizers. Here, we require that for any pair of an $\op X$- and a $\op Z$-boundary tile, they either do not intersect or that the intersection of their supports is entirely in the set of edges of the layout. Inspired by the rough and smooth boundaries of the surface code, we can for example choose the following sets of boundary stabilizers. 

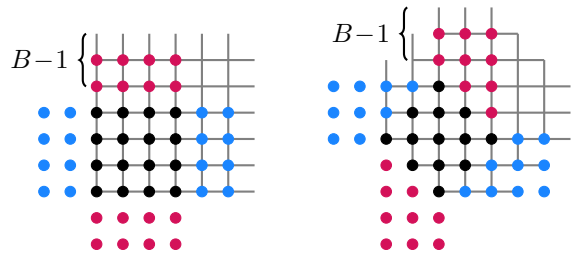
\begin{equation*}\label{eq: everythingwithbulkandboundary}
    \begin{tikzpicture}[scale = .7, baseline=(current bounding box.center)]
       \begin{scope}[xshift=0cm]
            \draw[decorate,decoration={brace},thick] (-0.2,2) -- (-0.2,3) node[midway,left=0.1cm]{\(B\!-\!1\)};
            \foreach \x in {0,...,5} {
                \draw[\latticecolor, \latticethickness] (\x/2,0) -- (\x/2,3);
            }
            
            \foreach \y in {0,...,5} {
                \draw[\latticecolor, \latticethickness] (0,\y/2) -- (3,\y/2);
            }
            
            \foreach \x in {0,...,3} {
                \foreach \y in {0,...,3} {
                    \draw[fill=black] (\x/2,\y/2) circle (0.1);
                }
            }
            
            \foreach \x in {0,...,3} {
                \draw[fill=good_red, color = good_red] (\x/2,-0.5) circle (0.1);
                \draw[fill=good_red, color = good_red] (\x/2,-1) circle (0.1);
                \draw[fill=good_red, color = good_red] (\x/2,2) circle (0.1);
                \draw[fill=good_red, color = good_red] (\x/2,2.5) circle (0.1);
            }
            
            \foreach \y in {0,...,3} {
                \draw[fill=good_blue, color = good_blue] (-0.5,\y/2) circle (0.1);
                \draw[fill=good_blue, color = good_blue] (-1,\y/2) circle (0.1);
                \draw[fill=good_blue, color = good_blue] (2,\y/2) circle (0.1);
                \draw[fill=good_blue, color = good_blue] (2.5,\y/2) circle (0.1);
            }
        \end{scope}
        \begin{scope}[xshift=6.5cm, yshift = 1cm]
            \draw[\latticecolor, \latticethickness] (-2/2, 0) -- (-2/2, 3/2);
            \draw[\latticecolor, \latticethickness] (-1/2, -1/2) -- (-1/2, 4/2);
            \draw[\latticecolor, \latticethickness] (0, -1) -- (0, 5/2);
            \draw[\latticecolor, \latticethickness] (1/2, -1) -- (1/2, 5/2);
            \draw[\latticecolor, \latticethickness] (1, -1) -- (1, 5/2);
            \draw[\latticecolor, \latticethickness] (3/2, -1/2) -- (3/2, 4/2);
            \draw[\latticecolor, \latticethickness] (2, 0) -- (2, 3/2);
            
            \draw[\latticecolor, \latticethickness] (0, -2/2) -- (3/2, -2/2);
            \draw[\latticecolor, \latticethickness] (-1/2, -1/2) -- (4/2, -1/2);
            \draw[\latticecolor, \latticethickness] (-2/2, -0/2) -- (5/2, -0/2);
            \draw[\latticecolor, \latticethickness] (-2/2, 1/2) -- (5/2, 1/2);
            \draw[\latticecolor, \latticethickness] (-2/2, 2/2) -- (5/2, 2/2);
            \draw[\latticecolor, \latticethickness] (-1/2, 3/2) -- (4/2, 3/2);
            \draw[\latticecolor, \latticethickness] (0/2, 4/2) -- (3/2, 4/2);
            
            \draw[fill=black] (-2/2,0/2) circle (0.1);
            \draw[fill=black] (-1/2,0/2) circle (0.1);
            \draw[fill=black] (-1/2,-1/2) circle (0.1);
            \draw[fill=black] (-1/2,1/2) circle (0.1);
            \draw[fill=black] (-0/2,-2/2) circle (0.1);
            \draw[fill=black] (-0/2,-1/2) circle (0.1);
            \draw[fill=black] (-0/2,0/2) circle (0.1);
            \draw[fill=black] (-0/2,1/2) circle (0.1);
            \draw[fill=black] (-0/2,2/2) circle (0.1);
            \draw[fill=black] (1/2,0/2) circle (0.1);
            \draw[fill=black] (1/2,-1/2) circle (0.1);
            \draw[fill=black] (1/2,1/2) circle (0.1);
            \draw[fill=black] (2/2, 0/2) circle (0.1);

            \draw[fill=good_red, color = good_red] (-2/2,-1/2) circle (0.1);
            \draw[fill=good_red, color = good_red] (-2/2,-2/2) circle (0.1);
            \draw[fill=good_red, color = good_red] (-2/2,-3/2) circle (0.1);
            \draw[fill=good_red, color = good_red] (-2/2,-4/2) circle (0.1);            \draw[fill=good_red, color = good_red] (-1/2,-2/2) circle (0.1);
            \draw[fill=good_red, color = good_red] (-1/2,-3/2) circle (0.1);
            \draw[fill=good_red, color = good_red] (-1/2,-4/2) circle (0.1);            \draw[fill=good_red, color = good_red] (0/2,-3/2) circle (0.1);
            \draw[fill=good_red, color = good_red] (0/2,-4/2) circle (0.1);
            
            \draw[fill=good_red, color = good_red] (0/2,3/2) circle (0.1);
            \draw[fill=good_red, color = good_red] (0/2,4/2) circle (0.1);   
            \draw[fill=good_red, color = good_red] (1/2,2/2) circle (0.1);
            \draw[fill=good_red, color = good_red] (1/2,3/2) circle (0.1);
            \draw[fill=good_red, color = good_red] (1/2,4/2) circle (0.1);     
            \draw[fill=good_red, color = good_red] (2/2,1/2) circle (0.1);
            \draw[fill=good_red, color = good_red] (2/2,2/2) circle (0.1);
            \draw[fill=good_red, color = good_red] (2/2,3/2) circle (0.1);
            \draw[fill=good_red, color = good_red] (2/2,4/2) circle (0.1);

            \draw[fill=good_blue, color = good_blue] (-4/2,2/2) circle (0.1);
            \draw[fill=good_blue, color = good_blue] (-3/2,2/2) circle (0.1);
            \draw[fill=good_blue, color = good_blue] (-2/2,2/2) circle (0.1);
            \draw[fill=good_blue, color = good_blue] (-1/2,2/2) circle (0.1);
            \draw[fill=good_blue, color = good_blue] (-4/2,1/2) circle (0.1);
            \draw[fill=good_blue, color = good_blue] (-3/2,1/2) circle (0.1);
            \draw[fill=good_blue, color = good_blue] (-2/2,1/2) circle (0.1);
            \draw[fill=good_blue, color = good_blue] (-4/2,0/2) circle (0.1);
            \draw[fill=good_blue, color = good_blue] (-3/2,0/2) circle (0.1);
            \draw[fill=good_blue, color = good_blue] (3/2,0/2) circle (0.1);
            \draw[fill=good_blue, color = good_blue] (4/2,0/2) circle (0.1);
            \draw[fill=good_blue, color = good_blue] (2/2,-1/2) circle (0.1);
            \draw[fill=good_blue, color = good_blue] (3/2,-1/2) circle (0.1);
            \draw[fill=good_blue, color = good_blue] (4/2,-1/2) circle (0.1);
            \draw[fill=good_blue, color = good_blue] (1/2,-2/2) circle (0.1);
            \draw[fill=good_blue, color = good_blue] (2/2,-2/2) circle (0.1);
            \draw[fill=good_blue, color = good_blue] (3/2,-2/2) circle (0.1);
            \draw[fill=good_blue, color = good_blue] (4/2,-2/2) circle (0.1);

            \draw[decorate,decoration={brace},thick] (-0.6,1.5) -- (-0.6,2.5) node[midway,left=0.1cm]{\(B\!-\!1\)};
        \end{scope}
    \end{tikzpicture}
\end{equation*}

It is clear that the choices of the sets of boundary stabilizers shown bove generalize to any sizes of the initial bulk stabilizer lattice as well as the boxes that contain the stabilizers. Note also that all boundary stabilizers commute. This is because any pair of boxes at a red and a blue dot does either not overlap or has their intersection of edges completely contained in the set of physical qubits, see \Cref{fig: masterfigure} (c) for a visualization.

We emphasize that so far, the construction is agnostic of the specific support of the stabilizer tiles. In fact, we only use that the support is contained in boxes of size $B \times B$, see \ref{it:condition for size tiles bb code}.

Finally, we fine-tune the layout to the specific support of the stabilizers. In more detail, we will first remove all qubits that are not supported in any $\op X$-type stabilizer or are not supported in any $\op Z$-type stabilizer. Finally, we remove all stabilizers whose support has become empty because of this. Note that this step largely depends on the supports of the stabilizers in the tiles.

\textit{Examples of tile codes ---} Despite the simplicity of this construction, we will now see that this approach can yield planar versions of BB codes that drastically outperform all known constructions in this direction \cite{eberhardt2024pruningqldpccodesbivariate, pecorari2024highrate}. 

We systematically searched for a variety of box sizes and stabilizer weights. We first estimated the distance of the codes using a probabilistic algorithm~\cite{GAP4, Pryadko2022} and then confirmed the results using an integer linear program solver.

The results of this search can be found in \Cref{tab:code we found}. We will now highlight a few concrete codes with attractive parameters.

\begin{table}
    \begin{tabular}{c|c|c|c|c}
         $B$\;& \;$w$\;& \;$[[n,k,d]]$\;& $kd^2/n$ \;&\;$\op X$- and $\op Z$-tiles\\
         \hline\hline
         3& 6& [[288,8,12]]& 4 & 
         \begin{tikzpicture}[scale=0.35, baseline= 8]
        \def\horizontalqubits{(0,0),(2,1), (2,2)}
        \def\verticalqubits{(0,2),(1,2),(2,0)}

        \def\horizontalqubitsz{(\boxsize - 1 - 0,\boxsize - 1 - 2),(\boxsize - 1 - 1,\boxsize - 1 - 2),(\boxsize - 1 - 2,\boxsize - 1 - 0)}
        \def\verticalqubitsz{(\boxsize - 1 - 0,\boxsize - 1 - 0),(\boxsize - 1 - 2,\boxsize - 1 - 1),(\boxsize - 1 - 2,\boxsize - 1 - 2)}
        \def\boxsize{3}
        
        \begin{scope}[xshift=0cm]
            \draw[\latticecolor, \latticethickness] (3,0) -- (0,0) -- (0,3) ;
            \draw[\latticecolor, \latticethickness] (1,0) -- (1,3);
            \draw[\latticecolor, \latticethickness] (2,0) -- (2,3);
            \draw[\latticecolor, \latticethickness] (0,1) -- (3,1);
            \draw[\latticecolor, \latticethickness] (0,2) -- (3,2);
            \fill[color = good_red, opacity = .0] (0,0) -- (3,0) -- (3,3) -- (0,3) -- cycle;
            \foreach \pos in \horizontalqubits {
                \draw[good_red, very thick] \pos -- ++(1,0);
            }
            \foreach \pos in \verticalqubits {
                \draw[good_red, very thick] \pos -- ++(0,1);
            }
            
            \draw[fill=good_red, color = good_red] (0,0) circle (0.15);
        \end{scope}

        \def\horizontalqubits{(0,0),(0,1), (2,2)}
        \def\verticalqubits{(0,2),(1,0),(2,0)}
        \begin{scope}[xshift=5cm]
            \draw[\latticecolor, \latticethickness] (3,0) -- (0,0) -- (0,3) ;
            \draw[\latticecolor, \latticethickness] (1,0) -- (1,3);
            \draw[\latticecolor, \latticethickness] (2,0) -- (2,3);
            \draw[\latticecolor, \latticethickness] (0,1) -- (3,1);
            \draw[\latticecolor, \latticethickness] (0,2) -- (3,2);
            
            \foreach \pos in \horizontalqubitsz {
                \draw[good_blue, very thick] \pos -- ++(1,0);
            }
            \foreach \pos in \verticalqubitsz {
                \draw[good_blue, very thick] \pos -- ++(0,1);
            }
            
            \draw[fill=good_blue, color = good_blue] (0,0) circle (0.15);
        \end{scope}
        \addvmargin{2mm}
    \end{tikzpicture}\\
         \hline
         3&8& [[288,8,14]]&$5.4$&    
         \begin{tikzpicture}[scale=0.35, baseline=8]
        \def\horizontalqubits{(0,0),(2,0), (0,1), (0,2)}
        \def\verticalqubits{(0,0),(0,2),(1,1), (2,2)}

        \def\horizontalqubitsz{
            (\boxsize - 1 - 0,\boxsize - 1 - 0),
            (\boxsize - 1 - 0,\boxsize - 1 - 2),
            (\boxsize - 1 - 1,\boxsize - 1 - 1),
            (\boxsize - 1 - 2,\boxsize - 1 - 2),
        }
        \def\verticalqubitsz{
            (\boxsize - 1 - 0,\boxsize - 1 - 0),
            (\boxsize - 1 - 2,\boxsize - 1 - 0),
            (\boxsize - 1 - 0,\boxsize - 1 - 1),
            (\boxsize - 1 - 0,\boxsize - 1 - 2),
        }
        \def\boxsize{3}
        
        \begin{scope}[xshift=0cm]
            \draw[\latticecolor, \latticethickness] (3,0) -- (0,0) -- (0,3) ;
            \draw[\latticecolor, \latticethickness] (1,0) -- (1,3);
            \draw[\latticecolor, \latticethickness] (2,0) -- (2,3);
            \draw[\latticecolor, \latticethickness] (0,1) -- (3,1);
            \draw[\latticecolor, \latticethickness] (0,2) -- (3,2);

            \foreach \pos in \horizontalqubits {
                \draw[good_red, very thick] \pos -- ++(1,0);
            }
            \foreach \pos in \verticalqubits {
                \draw[good_red, very thick] \pos -- ++(0,1);
            }
            
            \draw[fill=good_red, color = good_red] (0,0) circle (0.15);
        \end{scope}

        \def\horizontalqubits{(0,0),(0,1), (2,2)}
        \def\verticalqubits{(0,2),(1,0),(2,0)}
        \begin{scope}[xshift=5cm]
            \draw[\latticecolor, \latticethickness] (3,0) -- (0,0) -- (0,3) ;
            \draw[\latticecolor, \latticethickness] (1,0) -- (1,3);
            \draw[\latticecolor, \latticethickness] (2,0) -- (2,3);
            \draw[\latticecolor, \latticethickness] (0,1) -- (3,1);
            \draw[\latticecolor, \latticethickness] (0,2) -- (3,2);
            
            \foreach \pos in \horizontalqubitsz {
                \draw[good_blue, very thick] \pos -- ++(1,0);
            }
            \foreach \pos in \verticalqubitsz {
                \draw[good_blue, very thick] \pos -- ++(0,1);
            }
            
            \draw[fill=good_blue, color = good_blue] (0,0) circle (0.15);
        \end{scope}
                \addvmargin{1mm}

    \end{tikzpicture}
         
         \\
         \hline
         4& 8& [[288,18,13]]& 10.6 & 
    \begin{tikzpicture}[scale=0.35, baseline = 14]
        \def\horizontalqubits{(0, 0), (0, 3), (2, 2), (3, 0)}
        \def\verticalqubits{(0, 1), (1, 0), (1, 1), (3, 3)}

        \def\horizontalqubitsz{
            (\boxsize - 1 - 0,\boxsize - 1 - 1),
            (\boxsize - 1 - 1,\boxsize - 1 - 0),
            (\boxsize - 1 - 1,\boxsize - 1 - 1),
            (\boxsize - 1 - 3,\boxsize - 1 - 3)
        }
        \def\verticalqubitsz{
            (\boxsize - 1 - 0,\boxsize - 1 - 0),
            (\boxsize - 1 - 0,\boxsize - 1 - 3),
            (\boxsize - 1 - 2,\boxsize - 1 - 2),
            (\boxsize - 1 - 3,\boxsize - 1 - 0)            
            }
        \def\boxsize{4}
        
        \begin{scope}[xshift=0cm]
            \draw[\latticecolor, \latticethickness] (0,0) -- (4,0);
            \draw[\latticecolor, \latticethickness] (0,1) -- (4,1);
            \draw[\latticecolor, \latticethickness] (0,2) -- (4,2);
            \draw[\latticecolor, \latticethickness] (0,3) -- (4,3);
            \draw[\latticecolor, \latticethickness] (0,0) -- (0,4);
            \draw[\latticecolor, \latticethickness] (1,0) -- (1,4);
            \draw[\latticecolor, \latticethickness] (2,0) -- (2,4);
            \draw[\latticecolor, \latticethickness] (3,0) -- (3,4);

            \foreach \pos in \horizontalqubits {
                \draw[good_red, very thick] \pos -- ++(1,0);
            }
            \foreach \pos in \verticalqubits {
                \draw[good_red, very thick] \pos -- ++(0,1);
            }
            
            \draw[fill=good_red, color = good_red] (0,0) circle (0.15);
        \end{scope}

        \def\horizontalqubits{(0,0),(0,1), (2,2)}
        \def\verticalqubits{(0,2),(1,0),(2,0)}
        \begin{scope}[xshift=5cm]
            \draw[\latticecolor, \latticethickness] (0,0) -- (4,0);
            \draw[\latticecolor, \latticethickness] (0,1) -- (4,1);
            \draw[\latticecolor, \latticethickness] (0,2) -- (4,2);
            \draw[\latticecolor, \latticethickness] (0,3) -- (4,3);
            \draw[\latticecolor, \latticethickness] (0,0) -- (0,4);
            \draw[\latticecolor, \latticethickness] (1,0) -- (1,4);
            \draw[\latticecolor, \latticethickness] (2,0) -- (2,4);
            \draw[\latticecolor, \latticethickness] (3,0) -- (3,4);
            \foreach \pos in \horizontalqubitsz {
                \draw[good_blue, very thick] \pos -- ++(1,0);
            }
            \foreach \pos in \verticalqubitsz {
                \draw[good_blue, very thick] \pos -- ++(0,1);
            }
            
            \draw[fill=good_blue, color = good_blue] (0,0) circle (0.15);
        \end{scope}
                \addvmargin{1mm}

    \end{tikzpicture}
         \\
         \hline
         4 & 8 & [[512,18,19]]&  12.7 &
    \begin{tikzpicture}[scale=0.35, baseline = 14]
        \def\horizontalqubits{(0, 0), (0, 3), (2, 2), (3, 0)}
        \def\verticalqubits{(0, 1), (1, 0), (1, 1), (3, 3)}

        \def\horizontalqubitsz{
            (\boxsize - 1 - 0,\boxsize - 1 - 1),
            (\boxsize - 1 - 1,\boxsize - 1 - 0),
            (\boxsize - 1 - 1,\boxsize - 1 - 1),
            (\boxsize - 1 - 3,\boxsize - 1 - 3)
        }
        \def\verticalqubitsz{
            (\boxsize - 1 - 0,\boxsize - 1 - 0),
            (\boxsize - 1 - 0,\boxsize - 1 - 3),
            (\boxsize - 1 - 2,\boxsize - 1 - 2),
            (\boxsize - 1 - 3,\boxsize - 1 - 0)            
            }
        \def\boxsize{4}
        
        \begin{scope}[xshift=0cm]
            \draw[\latticecolor, \latticethickness] (0,0) -- (4,0);
            \draw[\latticecolor, \latticethickness] (0,1) -- (4,1);
            \draw[\latticecolor, \latticethickness] (0,2) -- (4,2);
            \draw[\latticecolor, \latticethickness] (0,3) -- (4,3);
            \draw[\latticecolor, \latticethickness] (0,0) -- (0,4);
            \draw[\latticecolor, \latticethickness] (1,0) -- (1,4);
            \draw[\latticecolor, \latticethickness] (2,0) -- (2,4);
            \draw[\latticecolor, \latticethickness] (3,0) -- (3,4);

            \foreach \pos in \horizontalqubits {
                \draw[good_red, very thick] \pos -- ++(1,0);
            }
            \foreach \pos in \verticalqubits {
                \draw[good_red, very thick] \pos -- ++(0,1);
            }
            
            \draw[fill=good_red, color = good_red] (0,0) circle (0.15);
        \end{scope}

        \def\horizontalqubits{(0,0),(0,1), (2,2)}
        \def\verticalqubits{(0,2),(1,0),(2,0)}
        \begin{scope}[xshift=5cm]
            \draw[\latticecolor, \latticethickness] (0,0) -- (4,0);
            \draw[\latticecolor, \latticethickness] (0,1) -- (4,1);
            \draw[\latticecolor, \latticethickness] (0,2) -- (4,2);
            \draw[\latticecolor, \latticethickness] (0,3) -- (4,3);
            \draw[\latticecolor, \latticethickness] (0,0) -- (0,4);
            \draw[\latticecolor, \latticethickness] (1,0) -- (1,4);
            \draw[\latticecolor, \latticethickness] (2,0) -- (2,4);
            \draw[\latticecolor, \latticethickness] (3,0) -- (3,4);
            \foreach \pos in \horizontalqubitsz {
                \draw[good_blue, very thick] \pos -- ++(1,0);
            }
            \foreach \pos in \verticalqubitsz {
                \draw[good_blue, very thick] \pos -- ++(0,1);
            }
            
            \draw[fill=good_blue, color = good_blue] (0,0) circle (0.15);
        \end{scope}
                \addvmargin{1mm}

    \end{tikzpicture}
    \end{tabular}
    \caption{Parameters of some tile codes. 
    We identified these examples by fixing the layout to the (unrotated) square layout and values for box size $B$, check weight $w$, and number of physical qubits $n$. 
    We then performed a brute force search among all possible tile configurations and selected those that produced codes with the highest efficiency ratio $kd^2/n$. 
    The distances presented are exact values, calculated using integer linear programming.
    }
    \label{tab:code we found}
\end{table}

We start by searching for codes using the unrotated layout using $10 \times 10$ copies of $\op X$- and $\op Z$-bulk tiles where we exhaustively check all stabilizers confined in a $3 \times 3$ box. Assuming that we will not remove qubits and stabilizers in the last step, the construction will always yield a code with 288 physical and 8 logical qubits. We find that the best possible distance is 12 giving a $[[288, 8, 12]]$ code. Note that this slight relaxation of locality therefore gives a 4-times improvement in $kd^2/n$ over the rotated surface code. We depict an example of stabilizer tiles giving these parameters here and refer to \Cref{sec:appendix weight 6} for more examples. 
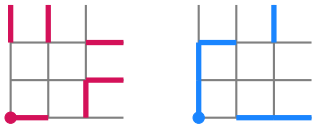
\begin{equation*}
    \begin{tikzpicture}[scale=0.5]
        \def\horizontalqubits{(0,0),(2,1), (2,2)}
        \def\verticalqubits{(0,2),(1,2),(2,0)}

        \def\horizontalqubitsz{(\boxsize - 1 - 0,\boxsize - 1 - 2),(\boxsize - 1 - 1,\boxsize - 1 - 2),(\boxsize - 1 - 2,\boxsize - 1 - 0)}
        \def\verticalqubitsz{(\boxsize - 1 - 0,\boxsize - 1 - 0),(\boxsize - 1 - 2,\boxsize - 1 - 1),(\boxsize - 1 - 2,\boxsize - 1 - 2)}
        \def\boxsize{3}
        
        \begin{scope}[xshift=0cm]
            \draw[\latticecolor, \latticethickness] (3,0) -- (0,0) -- (0,3) ;
            \draw[\latticecolor, \latticethickness] (1,0) -- (1,3);
            \draw[\latticecolor, \latticethickness] (2,0) -- (2,3);
            \draw[\latticecolor, \latticethickness] (0,1) -- (3,1);
            \draw[\latticecolor, \latticethickness] (0,2) -- (3,2);
            \fill[color = good_red, opacity = .0] (0,0) -- (3,0) -- (3,3) -- (0,3) -- cycle;
            \foreach \pos in \horizontalqubits {
                \draw[good_red, line width = \supportthickness] \pos -- ++(1,0);
            }
            \foreach \pos in \verticalqubits {
                \draw[good_red, line width = \supportthickness] \pos -- ++(0,1);
            }
            
            \draw[fill=good_red, color = good_red] (0,0) circle (0.15);
        \end{scope}

        \def\horizontalqubits{(0,0),(0,1), (2,2)}
        \def\verticalqubits{(0,2),(1,0),(2,0)}
        \begin{scope}[xshift=5cm]
            \draw[\latticecolor, \latticethickness] (3,0) -- (0,0) -- (0,3) ;
            \draw[\latticecolor, \latticethickness] (1,0) -- (1,3);
            \draw[\latticecolor, \latticethickness] (2,0) -- (2,3);
            \draw[\latticecolor, \latticethickness] (0,1) -- (3,1);
            \draw[\latticecolor, \latticethickness] (0,2) -- (3,2);
            
            \foreach \pos in \horizontalqubitsz {
                \draw[good_blue, line width = \supportthickness] \pos -- ++(1,0);
            }
            \foreach \pos in \verticalqubitsz {
                \draw[good_blue, line width = \supportthickness] \pos -- ++(0,1);
            }
            
            \draw[fill=good_blue, color = good_blue] (0,0) circle (0.15);
        \end{scope}
    \end{tikzpicture}
\end{equation*}
If we allow for weight 8 stabilizers instead of weight 6 in the same construction  stabilizers, we can even find codes with parameters $[[288, 8, 14]]$. The following is an example of stabilizer tiles realizing these parameters. 

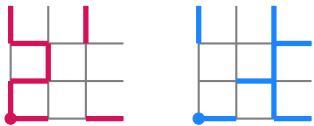
\begin{equation*}
    \begin{tikzpicture}[scale=0.5]
        \def\horizontalqubits{(0,0),(2,0), (0,1), (0,2)}
        \def\verticalqubits{(0,0),(0,2),(1,1), (2,2)}

        \def\horizontalqubitsz{
            (\boxsize - 1 - 0,\boxsize - 1 - 0),
            (\boxsize - 1 - 0,\boxsize - 1 - 2),
            (\boxsize - 1 - 1,\boxsize - 1 - 1),
            (\boxsize - 1 - 2,\boxsize - 1 - 2),
        }
        \def\verticalqubitsz{
            (\boxsize - 1 - 0,\boxsize - 1 - 0),
            (\boxsize - 1 - 2,\boxsize - 1 - 0),
            (\boxsize - 1 - 0,\boxsize - 1 - 1),
            (\boxsize - 1 - 0,\boxsize - 1 - 2),
        }
        \def\boxsize{3}
        
        \begin{scope}[xshift=0cm]
            \draw[\latticecolor, \latticethickness] (3,0) -- (0,0) -- (0,3) ;
            \draw[\latticecolor, \latticethickness] (1,0) -- (1,3);
            \draw[\latticecolor, \latticethickness] (2,0) -- (2,3);
            \draw[\latticecolor, \latticethickness] (0,1) -- (3,1);
            \draw[\latticecolor, \latticethickness] (0,2) -- (3,2);

            \foreach \pos in \horizontalqubits {
                \draw[good_red, line width = \supportthickness] \pos -- ++(1,0);
            }
            \foreach \pos in \verticalqubits {
                \draw[good_red, line width = \supportthickness] \pos -- ++(0,1);
            }
            
            \draw[fill=good_red, color = good_red] (0,0) circle (0.15);
        \end{scope}

        \def\horizontalqubits{(0,0),(0,1), (2,2)}
        \def\verticalqubits{(0,2),(1,0),(2,0)}
        \begin{scope}[xshift=5cm]
            \draw[\latticecolor, \latticethickness] (3,0) -- (0,0) -- (0,3) ;
            \draw[\latticecolor, \latticethickness] (1,0) -- (1,3);
            \draw[\latticecolor, \latticethickness] (2,0) -- (2,3);
            \draw[\latticecolor, \latticethickness] (0,1) -- (3,1);
            \draw[\latticecolor, \latticethickness] (0,2) -- (3,2);
            
            \foreach \pos in \horizontalqubitsz {
                \draw[good_blue,  line width = \supportthickness] \pos -- ++(1,0);
            }
            \foreach \pos in \verticalqubitsz {
                \draw[good_blue, line width = \supportthickness] \pos -- ++(0,1);
            }
            
            \draw[fill=good_blue, color = good_blue] (0,0) circle (0.15);
        \end{scope}
    \end{tikzpicture}
\end{equation*}
\noindent Refer to \Cref{sec: appendix weight 8} for more examples.

Finally, if we allow for weight 8 stabilizer tiles that are confined in $4 \times 4$-boxes, we can obtain codes with parameters $[[288, 18, 13]]$ and [[512,18,19]] using the unrotated square layouts with $9 \times 9$ and $13 \times 13$ copies of the bulk tiles, respectively. 
They give a 10- to 12-times improvement over the rotated surface code. We depict examples of stabilizer tiles here. 
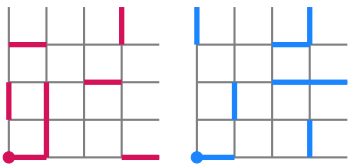
\begin{equation*}
    \begin{tikzpicture}[scale=0.5]
        \def\horizontalqubits{(0, 0), (0, 3), (2, 2), (3, 0)}
        \def\verticalqubits{(0, 1), (1, 0), (1, 1), (3, 3)}

        \def\horizontalqubitsz{
            (\boxsize - 1 - 0,\boxsize - 1 - 1),
            (\boxsize - 1 - 1,\boxsize - 1 - 0),
            (\boxsize - 1 - 1,\boxsize - 1 - 1),
            (\boxsize - 1 - 3,\boxsize - 1 - 3)
        }
        \def\verticalqubitsz{
            (\boxsize - 1 - 0,\boxsize - 1 - 0),
            (\boxsize - 1 - 0,\boxsize - 1 - 3),
            (\boxsize - 1 - 2,\boxsize - 1 - 2),
            (\boxsize - 1 - 3,\boxsize - 1 - 0)            
            }
        \def\boxsize{4}
        
        \begin{scope}[xshift=0cm]
            \draw[\latticecolor, \latticethickness] (0,0) -- (4,0);
            \draw[\latticecolor, \latticethickness] (0,1) -- (4,1);
            \draw[\latticecolor, \latticethickness] (0,2) -- (4,2);
            \draw[\latticecolor, \latticethickness] (0,3) -- (4,3);
            \draw[\latticecolor, \latticethickness] (0,0) -- (0,4);
            \draw[\latticecolor, \latticethickness] (1,0) -- (1,4);
            \draw[\latticecolor, \latticethickness] (2,0) -- (2,4);
            \draw[\latticecolor, \latticethickness] (3,0) -- (3,4);

            \foreach \pos in \horizontalqubits {
                \draw[good_red, line width=2pt] \pos -- ++(1,0);
            }
            \foreach \pos in \verticalqubits {
                \draw[good_red, line width=2pt] \pos -- ++(0,1);
            }
            
            \draw[fill=good_red, color = good_red] (0,0) circle (0.15);
        \end{scope}

        \def\horizontalqubits{(0,0),(0,1), (2,2)}
        \def\verticalqubits{(0,2),(1,0),(2,0)}
        \begin{scope}[xshift=5cm]
            \draw[\latticecolor, \latticethickness] (0,0) -- (4,0);
            \draw[\latticecolor, \latticethickness] (0,1) -- (4,1);
            \draw[\latticecolor, \latticethickness] (0,2) -- (4,2);
            \draw[\latticecolor, \latticethickness] (0,3) -- (4,3);
            \draw[\latticecolor, \latticethickness] (0,0) -- (0,4);
            \draw[\latticecolor, \latticethickness] (1,0) -- (1,4);
            \draw[\latticecolor, \latticethickness] (2,0) -- (2,4);
            \draw[\latticecolor, \latticethickness] (3,0) -- (3,4);
            \foreach \pos in \horizontalqubitsz {
                \draw[good_blue, line width=2pt] \pos -- ++(1,0);
            }
            \foreach \pos in \verticalqubitsz {
                \draw[good_blue, line width=2pt] \pos -- ++(0,1);
            }
            
            \draw[fill=good_blue, color = good_blue] (0,0) circle (0.15);
        \end{scope}
    \end{tikzpicture}
\end{equation*}

For higher stabilizer weights, we did not perform an exhaustive search. By a randomized search, for weight 10 and $4\times 4$ boxes we found a 
$[[512, 18,\leq\!23]]$-code with the following stabilizer tiles. We strongly expect that $d=23$ and hence $kd^2/n\approx 18.6$ though we did not formally verify this using integer linear programming.
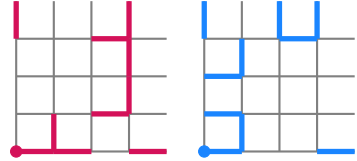
\begin{equation*}
    \begin{tikzpicture}[scale=0.5]
        \def\horizontalqubits{(0, 0), (1, 0), (2, 1), (2, 3), (3, 0)}
        \def\verticalqubits{(0, 3), (1, 0), (3, 1), (3, 2), (3, 3)}

        \def\horizontalqubitsz{ 
            (\boxsize - 1 - 0,\boxsize - 1 - 3), 
            (\boxsize - 1 - 1,\boxsize - 1 - 0), 
            (\boxsize - 1 - 3,\boxsize - 1 - 1), 
            (\boxsize - 1 - 3,\boxsize - 1 - 2), 
            (\boxsize - 1 - 3,\boxsize - 1 - 3)  
        }
        \def\verticalqubitsz{ 
            (\boxsize - 1 - 0,\boxsize - 1 - 0), 
            (\boxsize - 1 - 1,\boxsize - 1 - 0), 
            (\boxsize - 1 - 2,\boxsize - 1 - 1), 
            (\boxsize - 1 - 2,\boxsize - 1 - 3), 
            (\boxsize - 1 - 3,\boxsize - 1 - 0)  
            }
        \def\boxsize{4}
        
        \begin{scope}[xshift=0cm]
            \draw[\latticecolor, \latticethickness] (0,0) -- (4,0);
            \draw[\latticecolor, \latticethickness] (0,1) -- (4,1);
            \draw[\latticecolor, \latticethickness] (0,2) -- (4,2);
            \draw[\latticecolor, \latticethickness] (0,3) -- (4,3);
            \draw[\latticecolor, \latticethickness] (0,0) -- (0,4);
            \draw[\latticecolor, \latticethickness] (1,0) -- (1,4);
            \draw[\latticecolor, \latticethickness] (2,0) -- (2,4);
            \draw[\latticecolor, \latticethickness] (3,0) -- (3,4);

            \foreach \pos in \horizontalqubits {
                \draw[good_red, line width=2pt] \pos -- ++(1,0);
            }
            \foreach \pos in \verticalqubits {
                \draw[good_red, line width=2pt] \pos -- ++(0,1);
            }
            
            \draw[fill=good_red, color = good_red] (0,0) circle (0.15);
        \end{scope}
        \begin{scope}[xshift=5cm]
            \draw[\latticecolor, \latticethickness] (0,0) -- (4,0);
            \draw[\latticecolor, \latticethickness] (0,1) -- (4,1);
            \draw[\latticecolor, \latticethickness] (0,2) -- (4,2);
            \draw[\latticecolor, \latticethickness] (0,3) -- (4,3);
            \draw[\latticecolor, \latticethickness] (0,0) -- (0,4);
            \draw[\latticecolor, \latticethickness] (1,0) -- (1,4);
            \draw[\latticecolor, \latticethickness] (2,0) -- (2,4);
            \draw[\latticecolor, \latticethickness] (3,0) -- (3,4);
            \foreach \pos in \horizontalqubitsz {
                \draw[good_blue, line width=2pt] \pos -- ++(1,0);
            }
            \foreach \pos in \verticalqubitsz {
                \draw[good_blue, line width=2pt] \pos -- ++(0,1);
            }
            
            \draw[fill=good_blue, color = good_blue] (0,0) circle (0.15);
        \end{scope}
    \end{tikzpicture}
\end{equation*}

We also mention that all results from \cite{eberhardt2024pruningqldpccodesbivariate} on pruning of hypergraph product codes of cyclic codes can be reconstructed from our construction using the unrotated bulk stabilizer lattice. For a proof we refer to \Cref{sec: appendix hgp}. 
In addition, we remark that the rotated surface code can be recovered from our construction using a rotated layout. 
We also found some codes with weight 6 or 8 stabilizers using the rotated layout, but their parameters were not as good as those using the unrotated layout.

\textit{Discussion ---} 
In this Letter, we give a simple but powerful construction for 2D-local codes on a planar lattice. With this, we address an important problem in making qLDPC codes practical for near- and mid-term superconducting hardware. By relaxing the locality condition of the surface code slightly, we observe up to 12 times improvement over the rotated surface code and find codes that outperform current state-of-the-art proposals for the implementation of qLDPC codes.

Our construction is simple and general, and we believe that there are many ways forward to construct further tile codes; e.g.\ using other bulk stabilizer tile layouts like rotations and layouts with holes or polygonal boundaries. Another future line of research is to investigate protocols and design ancillary systems for performing logical computation on the codes we found.
We expect that one can adapt the QEC cycle for BB codes developed in \cite{bravyi2023highthreshold} to tile codes. It would be very interesting to see how the boundary impacts the numerical simulations and threshold.
Furthermore, one could explore the construction of non-CSS versions of tile codes by applying Clifford deformations similar to the XZZX surface code.

\textit{Acknowledgements ---} We thank Yu-An Chen and Zhijian Liang for fruitful discussions. JNE was supported by Deutsche Forschungsgemeinschaft (DFG),
project number 45744154, Equivariant K-motives and Koszul duality.

\bibliographystyle{apsrev4-1}
\bibliography{main}

\begin{thebibliography}{28}%
\makeatletter
\providecommand \@ifxundefined [1]{%
 \@ifx{#1\undefined}
}%
\providecommand \@ifnum [1]{%
 \ifnum #1\expandafter \@firstoftwo
 \else \expandafter \@secondoftwo
 \fi
}%
\providecommand \@ifx [1]{%
 \ifx #1\expandafter \@firstoftwo
 \else \expandafter \@secondoftwo
 \fi
}%
\providecommand \natexlab [1]{#1}%
\providecommand \enquote  [1]{``#1''}%
\providecommand \bibnamefont  [1]{#1}%
\providecommand \bibfnamefont [1]{#1}%
\providecommand \citenamefont [1]{#1}%
\providecommand \href@noop [0]{\@secondoftwo}%
\providecommand \href [0]{\begingroup \@sanitize@url \@href}%
\providecommand \@href[1]{\@@startlink{#1}\@@href}%
\providecommand \@@href[1]{\endgroup#1\@@endlink}%
\providecommand \@sanitize@url [0]{\catcode `\\12\catcode `\$12\catcode `\&12\catcode `\#12\catcode `\^12\catcode `\_12\catcode `\%12\relax}%
\providecommand \@@startlink[1]{}%
\providecommand \@@endlink[0]{}%
\providecommand \url  [0]{\begingroup\@sanitize@url \@url }%
\providecommand \@url [1]{\endgroup\@href {#1}{\urlprefix }}%
\providecommand \urlprefix  [0]{URL }%
\providecommand \Eprint [0]{\href }%
\providecommand \doibase [0]{http://dx.doi.org/}%
\providecommand \selectlanguage [0]{\@gobble}%
\providecommand \bibinfo  [0]{\@secondoftwo}%
\providecommand \bibfield  [0]{\@secondoftwo}%
\providecommand \translation [1]{[#1]}%
\providecommand \BibitemOpen [0]{}%
\providecommand \bibitemStop [0]{}%
\providecommand \bibitemNoStop [0]{.\EOS\space}%
\providecommand \EOS [0]{\spacefactor3000\relax}%
\providecommand \BibitemShut  [1]{\csname bibitem#1\endcsname}%
\let\auto@bib@innerbib\@empty
\bibitem [{\citenamefont {Bravyi}\ and\ \citenamefont {Kitaev}(1998)}]{bravyi1998quantum}%
  \BibitemOpen
  \bibfield  {author} {\bibinfo {author} {\bibfnamefont {S.~B.}\ \bibnamefont {Bravyi}}\ and\ \bibinfo {author} {\bibfnamefont {A.~Y.}\ \bibnamefont {Kitaev}},\ }\href@noop {} {\enquote {\bibinfo {title} {Quantum codes on a lattice with boundary},}\ } (\bibinfo {year} {1998}),\ \Eprint {http://arxiv.org/abs/quant-ph/9811052} {arXiv:quant-ph/9811052 [quant-ph]} \BibitemShut {NoStop}%
\bibitem [{\citenamefont {Freedman}\ and\ \citenamefont {Meyer}(2001)}]{freedman2001projective}%
  \BibitemOpen
  \bibfield  {author} {\bibinfo {author} {\bibfnamefont {M.~H.}\ \bibnamefont {Freedman}}\ and\ \bibinfo {author} {\bibfnamefont {D.~A.}\ \bibnamefont {Meyer}},\ }\href@noop {} {\bibfield  {journal} {\bibinfo  {journal} {Foundations of Computational Mathematics}\ }\textbf {\bibinfo {volume} {1}},\ \bibinfo {pages} {325} (\bibinfo {year} {2001})}\BibitemShut {NoStop}%
\bibitem [{\citenamefont {Andersen}\ \emph {et~al.}(2020)\citenamefont {Andersen}, \citenamefont {Remm}, \citenamefont {Lazar}, \citenamefont {Krinner}, \citenamefont {Lacroix}, \citenamefont {Norris}, \citenamefont {Gabureac}, \citenamefont {Eichler},\ and\ \citenamefont {Wallraff}}]{Andersen_2020}%
  \BibitemOpen
  \bibfield  {author} {\bibinfo {author} {\bibfnamefont {C.~K.}\ \bibnamefont {Andersen}}, \bibinfo {author} {\bibfnamefont {A.}~\bibnamefont {Remm}}, \bibinfo {author} {\bibfnamefont {S.}~\bibnamefont {Lazar}}, \bibinfo {author} {\bibfnamefont {S.}~\bibnamefont {Krinner}}, \bibinfo {author} {\bibfnamefont {N.}~\bibnamefont {Lacroix}}, \bibinfo {author} {\bibfnamefont {G.~J.}\ \bibnamefont {Norris}}, \bibinfo {author} {\bibfnamefont {M.}~\bibnamefont {Gabureac}}, \bibinfo {author} {\bibfnamefont {C.}~\bibnamefont {Eichler}}, \ and\ \bibinfo {author} {\bibfnamefont {A.}~\bibnamefont {Wallraff}},\ }\href {\doibase 10.1038/s41567-020-0920-y} {\bibfield  {journal} {\bibinfo  {journal} {Nature Physics}\ }\textbf {\bibinfo {volume} {16}},\ \bibinfo {pages} {875–880} (\bibinfo {year} {2020})}\BibitemShut {NoStop}%
\bibitem [{\citenamefont {Krinner}\ \emph {et~al.}(2022)\citenamefont {Krinner}, \citenamefont {Lacroix}, \citenamefont {Remm}, \citenamefont {Di~Paolo}, \citenamefont {Genois}, \citenamefont {Leroux}, \citenamefont {Hellings}, \citenamefont {Lazar}, \citenamefont {Swiadek}, \citenamefont {Herrmann}, \citenamefont {Norris}, \citenamefont {Andersen}, \citenamefont {Müller}, \citenamefont {Blais}, \citenamefont {Eichler},\ and\ \citenamefont {Wallraff}}]{Krinner_2022}%
  \BibitemOpen
  \bibfield  {author} {\bibinfo {author} {\bibfnamefont {S.}~\bibnamefont {Krinner}}, \bibinfo {author} {\bibfnamefont {N.}~\bibnamefont {Lacroix}}, \bibinfo {author} {\bibfnamefont {A.}~\bibnamefont {Remm}}, \bibinfo {author} {\bibfnamefont {A.}~\bibnamefont {Di~Paolo}}, \bibinfo {author} {\bibfnamefont {E.}~\bibnamefont {Genois}}, \bibinfo {author} {\bibfnamefont {C.}~\bibnamefont {Leroux}}, \bibinfo {author} {\bibfnamefont {C.}~\bibnamefont {Hellings}}, \bibinfo {author} {\bibfnamefont {S.}~\bibnamefont {Lazar}}, \bibinfo {author} {\bibfnamefont {F.}~\bibnamefont {Swiadek}}, \bibinfo {author} {\bibfnamefont {J.}~\bibnamefont {Herrmann}}, \bibinfo {author} {\bibfnamefont {G.~J.}\ \bibnamefont {Norris}}, \bibinfo {author} {\bibfnamefont {C.~K.}\ \bibnamefont {Andersen}}, \bibinfo {author} {\bibfnamefont {M.}~\bibnamefont {Müller}}, \bibinfo {author} {\bibfnamefont {A.}~\bibnamefont {Blais}}, \bibinfo {author} {\bibfnamefont {C.}~\bibnamefont {Eichler}}, \ and\ \bibinfo {author} {\bibfnamefont {A.}~\bibnamefont {Wallraff}},\ }\href {\doibase 10.1038/s41586-022-04566-8} {\bibfield  {journal} {\bibinfo  {journal} {Nature}\ }\textbf {\bibinfo {volume} {605}},\ \bibinfo {pages} {669–674} (\bibinfo {year} {2022})}\BibitemShut {NoStop}%
\bibitem [{\citenamefont {{Google Quantum AI}}\ and\ \citenamefont {collaborators}(2023)}]{google2023}%
  \BibitemOpen
  \bibfield  {author} {\bibinfo {author} {\bibnamefont {{Google Quantum AI}}}\ and\ \bibinfo {author} {\bibnamefont {collaborators}},\ }\href {\doibase 10.1038/s41586-022-05434-1} {\bibfield  {journal} {\bibinfo  {journal} {Nature}\ }\textbf {\bibinfo {volume} {614}},\ \bibinfo {pages} {676–681} (\bibinfo {year} {2023})}\BibitemShut {NoStop}%
\bibitem [{\citenamefont {{Google Quantum AI}}\ and\ \citenamefont {Collaborators}(2024)}]{google2024}%
  \BibitemOpen
  \bibfield  {author} {\bibinfo {author} {\bibnamefont {{Google Quantum AI}}}\ and\ \bibinfo {author} {\bibnamefont {Collaborators}},\ }\href {\doibase 10.1038/s41586-024-08449-y} {\bibfield  {journal} {\bibinfo  {journal} {Nature}\ }\textbf {\bibinfo {volume} {638}},\ \bibinfo {pages} {920–926} (\bibinfo {year} {2024})}\BibitemShut {NoStop}%
\bibitem [{\citenamefont {Breuckmann}\ and\ \citenamefont {Eberhardt}(2021)}]{breuckmannQuantumLowDensityParityCheck2021a}%
  \BibitemOpen
  \bibfield  {author} {\bibinfo {author} {\bibfnamefont {N.~P.}\ \bibnamefont {Breuckmann}}\ and\ \bibinfo {author} {\bibfnamefont {J.~N.}\ \bibnamefont {Eberhardt}},\ }\href {\doibase 10.1103/PRXQuantum.2.040101} {\bibfield  {journal} {\bibinfo  {journal} {PRX Quantum}\ }\textbf {\bibinfo {volume} {2}},\ \bibinfo {pages} {040101} (\bibinfo {year} {2021})},\ \Eprint {http://arxiv.org/abs/2103.06309} {arxiv:2103.06309 [quant-ph]} \BibitemShut {NoStop}%
\bibitem [{\citenamefont {Bravyi}\ \emph {et~al.}(2010)\citenamefont {Bravyi}, \citenamefont {Poulin},\ and\ \citenamefont {Terhal}}]{Bravyi_2010}%
  \BibitemOpen
  \bibfield  {author} {\bibinfo {author} {\bibfnamefont {S.}~\bibnamefont {Bravyi}}, \bibinfo {author} {\bibfnamefont {D.}~\bibnamefont {Poulin}}, \ and\ \bibinfo {author} {\bibfnamefont {B.}~\bibnamefont {Terhal}},\ }\href {\doibase 10.1103/physrevlett.104.050503} {\bibfield  {journal} {\bibinfo  {journal} {Physical Review Letters}\ }\textbf {\bibinfo {volume} {104}} (\bibinfo {year} {2010}),\ 10.1103/physrevlett.104.050503}\BibitemShut {NoStop}%
\bibitem [{\citenamefont {Vigneau}\ \emph {et~al.}(2025)\citenamefont {Vigneau}, \citenamefont {Majumder}, \citenamefont {Rath}, \citenamefont {Parrado-Rodríguez}, \citenamefont {Pereira}, \citenamefont {Pogorzalek}, \citenamefont {Jones}, \citenamefont {Wurz}, \citenamefont {Renger}, \citenamefont {Verjauw}, \citenamefont {Yang}, \citenamefont {Ku}, \citenamefont {Kindel}, \citenamefont {Deppe},\ and\ \citenamefont {Heinsoo}}]{vigneau2025quantumerrordetectionqubitresonator}%
  \BibitemOpen
  \bibfield  {author} {\bibinfo {author} {\bibfnamefont {F.}~\bibnamefont {Vigneau}}, \bibinfo {author} {\bibfnamefont {S.}~\bibnamefont {Majumder}}, \bibinfo {author} {\bibfnamefont {A.}~\bibnamefont {Rath}}, \bibinfo {author} {\bibfnamefont {P.}~\bibnamefont {Parrado-Rodríguez}}, \bibinfo {author} {\bibfnamefont {F.~R.~F.}\ \bibnamefont {Pereira}}, \bibinfo {author} {\bibfnamefont {S.}~\bibnamefont {Pogorzalek}}, \bibinfo {author} {\bibfnamefont {T.}~\bibnamefont {Jones}}, \bibinfo {author} {\bibfnamefont {N.}~\bibnamefont {Wurz}}, \bibinfo {author} {\bibfnamefont {M.}~\bibnamefont {Renger}}, \bibinfo {author} {\bibfnamefont {J.}~\bibnamefont {Verjauw}}, \bibinfo {author} {\bibfnamefont {P.}~\bibnamefont {Yang}}, \bibinfo {author} {\bibfnamefont {H.-S.}\ \bibnamefont {Ku}}, \bibinfo {author} {\bibfnamefont {W.}~\bibnamefont {Kindel}}, \bibinfo {author} {\bibfnamefont {F.}~\bibnamefont {Deppe}}, \ and\ \bibinfo {author} {\bibfnamefont {J.}~\bibnamefont {Heinsoo}},\ }\href {https://arxiv.org/abs/2503.12869} {\enquote {\bibinfo {title} {Quantum error detection in qubit-resonator star architecture},}\ } (\bibinfo {year} {2025}),\ \Eprint {http://arxiv.org/abs/2503.12869} {arXiv:2503.12869 [quant-ph]} \BibitemShut {NoStop}%
\bibitem [{\citenamefont {Renger}\ \emph {et~al.}(2025)\citenamefont {Renger}, \citenamefont {Verjauw}, \citenamefont {Wurz}, \citenamefont {Hosseinkhani}, \citenamefont {Ockeloen-Korppi}, \citenamefont {Liu}, \citenamefont {Rath}, \citenamefont {Thapa}, \citenamefont {Vigneau}, \citenamefont {Wybo}, \citenamefont {Bergholm}, \citenamefont {Chan}, \citenamefont {Csatári}, \citenamefont {Dahl}, \citenamefont {Davletkaliyev}, \citenamefont {Giri}, \citenamefont {Gusenkova}, \citenamefont {Heimonen}, \citenamefont {Hiltunen}, \citenamefont {Hsu}, \citenamefont {Hyyppä}, \citenamefont {Ikonen}, \citenamefont {Jones}, \citenamefont {Khalid}, \citenamefont {Kim}, \citenamefont {Koistinen}, \citenamefont {Komlev}, \citenamefont {Kotilahti}, \citenamefont {Kukushkin}, \citenamefont {Lamprich}, \citenamefont {Landra}, \citenamefont {Lee}, \citenamefont {Li}, \citenamefont {Liebermann}, \citenamefont {Majumder}, \citenamefont {Mäntylä}, \citenamefont {Marxer}, \citenamefont {van~de Griend}, \citenamefont {Milchakov}, \citenamefont {Mrożek}, \citenamefont {Nath}, \citenamefont {Orell}, \citenamefont {Papič}, \citenamefont {Partanen}, \citenamefont {Plyushch}, \citenamefont {Pogorzalek}, \citenamefont {Ritvas}, \citenamefont {Romero}, \citenamefont {Sampo}, \citenamefont {Seppälä}, \citenamefont {Selinmaa}, \citenamefont {Sundström}, \citenamefont {Takmakov}, \citenamefont {Tarasinski}, \citenamefont {Tuorila}, \citenamefont {Tyrkkö}, \citenamefont {Välimaa}, \citenamefont {Wesdorp}, \citenamefont {Yang}, \citenamefont {Yu}, \citenamefont {Heinsoo}, \citenamefont {Vepsäläinen}, \citenamefont {Kindel}, \citenamefont {Ku},\ and\ \citenamefont {Deppe}}]{renger2025superconductingqubitresonatorquantumprocessor}%
  \BibitemOpen
  \bibfield  {author} {\bibinfo {author} {\bibfnamefont {M.}~\bibnamefont {Renger}}, \bibinfo {author} {\bibfnamefont {J.}~\bibnamefont {Verjauw}}, \bibinfo {author} {\bibfnamefont {N.}~\bibnamefont {Wurz}}, \bibinfo {author} {\bibfnamefont {A.}~\bibnamefont {Hosseinkhani}}, \bibinfo {author} {\bibfnamefont {C.}~\bibnamefont {Ockeloen-Korppi}}, \bibinfo {author} {\bibfnamefont {W.}~\bibnamefont {Liu}}, \bibinfo {author} {\bibfnamefont {A.}~\bibnamefont {Rath}}, \bibinfo {author} {\bibfnamefont {M.~J.}\ \bibnamefont {Thapa}}, \bibinfo {author} {\bibfnamefont {F.}~\bibnamefont {Vigneau}}, \bibinfo {author} {\bibfnamefont {E.}~\bibnamefont {Wybo}}, \bibinfo {author} {\bibfnamefont {V.}~\bibnamefont {Bergholm}}, \bibinfo {author} {\bibfnamefont {C.~F.}\ \bibnamefont {Chan}}, \bibinfo {author} {\bibfnamefont {B.}~\bibnamefont {Csatári}}, \bibinfo {author} {\bibfnamefont {S.}~\bibnamefont {Dahl}}, \bibinfo {author} {\bibfnamefont {R.}~\bibnamefont {Davletkaliyev}}, \bibinfo {author} {\bibfnamefont {R.}~\bibnamefont {Giri}}, \bibinfo {author} {\bibfnamefont {D.}~\bibnamefont {Gusenkova}}, \bibinfo {author} {\bibfnamefont {H.}~\bibnamefont {Heimonen}}, \bibinfo {author} {\bibfnamefont {T.}~\bibnamefont {Hiltunen}}, \bibinfo {author} {\bibfnamefont {H.}~\bibnamefont {Hsu}}, \bibinfo {author} {\bibfnamefont {E.}~\bibnamefont {Hyyppä}}, \bibinfo {author} {\bibfnamefont {J.}~\bibnamefont {Ikonen}}, \bibinfo {author} {\bibfnamefont {T.}~\bibnamefont {Jones}}, \bibinfo {author} {\bibfnamefont {S.}~\bibnamefont {Khalid}}, \bibinfo {author} {\bibfnamefont {S.-G.}\ \bibnamefont {Kim}}, \bibinfo {author} {\bibfnamefont {M.}~\bibnamefont {Koistinen}}, \bibinfo {author} {\bibfnamefont {A.}~\bibnamefont {Komlev}}, \bibinfo {author} {\bibfnamefont {J.}~\bibnamefont {Kotilahti}}, \bibinfo {author} {\bibfnamefont {V.}~\bibnamefont {Kukushkin}}, \bibinfo {author} {\bibfnamefont {J.}~\bibnamefont {Lamprich}}, \bibinfo {author} {\bibfnamefont {A.}~\bibnamefont {Landra}}, \bibinfo {author} {\bibfnamefont {L.-H.}\ \bibnamefont {Lee}}, \bibinfo {author} {\bibfnamefont {T.}~\bibnamefont {Li}}, \bibinfo {author} {\bibfnamefont {P.}~\bibnamefont {Liebermann}}, \bibinfo {author} {\bibfnamefont {S.}~\bibnamefont {Majumder}}, \bibinfo {author} {\bibfnamefont {J.}~\bibnamefont {Mäntylä}}, \bibinfo {author} {\bibfnamefont {F.}~\bibnamefont {Marxer}}, \bibinfo {author} {\bibfnamefont {A.~M.}\ \bibnamefont {van~de Griend}}, \bibinfo {author} {\bibfnamefont {V.}~\bibnamefont {Milchakov}}, \bibinfo {author} {\bibfnamefont {J.}~\bibnamefont {Mrożek}}, \bibinfo {author} {\bibfnamefont {J.}~\bibnamefont {Nath}}, \bibinfo {author} {\bibfnamefont {T.}~\bibnamefont {Orell}}, \bibinfo {author} {\bibfnamefont {M.}~\bibnamefont {Papič}}, \bibinfo {author} {\bibfnamefont {M.}~\bibnamefont {Partanen}}, \bibinfo {author} {\bibfnamefont {A.}~\bibnamefont {Plyushch}}, \bibinfo {author} {\bibfnamefont {S.}~\bibnamefont {Pogorzalek}}, \bibinfo {author} {\bibfnamefont {J.}~\bibnamefont {Ritvas}}, \bibinfo {author} {\bibfnamefont {P.~F.}\ \bibnamefont {Romero}}, \bibinfo {author} {\bibfnamefont {V.}~\bibnamefont {Sampo}}, \bibinfo {author} {\bibfnamefont {M.}~\bibnamefont {Seppälä}}, \bibinfo {author} {\bibfnamefont {V.}~\bibnamefont {Selinmaa}}, \bibinfo {author} {\bibfnamefont {L.}~\bibnamefont {Sundström}}, \bibinfo {author} {\bibfnamefont {I.}~\bibnamefont {Takmakov}}, \bibinfo {author} {\bibfnamefont {B.}~\bibnamefont {Tarasinski}}, \bibinfo {author} {\bibfnamefont {J.}~\bibnamefont {Tuorila}}, \bibinfo {author} {\bibfnamefont {O.}~\bibnamefont {Tyrkkö}}, \bibinfo {author} {\bibfnamefont {A.}~\bibnamefont {Välimaa}}, \bibinfo {author} {\bibfnamefont {J.}~\bibnamefont {Wesdorp}}, \bibinfo {author} {\bibfnamefont {P.}~\bibnamefont {Yang}}, \bibinfo {author} {\bibfnamefont {L.}~\bibnamefont {Yu}}, \bibinfo {author} {\bibfnamefont {J.}~\bibnamefont {Heinsoo}}, \bibinfo {author} {\bibfnamefont {A.}~\bibnamefont {Vepsäläinen}}, \bibinfo {author} {\bibfnamefont {W.}~\bibnamefont {Kindel}}, \bibinfo {author} {\bibfnamefont {H.-S.}\ \bibnamefont {Ku}}, \ and\ \bibinfo {author} {\bibfnamefont {F.}~\bibnamefont {Deppe}},\ }\href {https://arxiv.org/abs/2503.10903} {\enquote {\bibinfo {title} {A superconducting qubit-resonator quantum processor with effective all-to-all connectivity},}\ } (\bibinfo {year} {2025}),\ \Eprint {http://arxiv.org/abs/2503.10903} {arXiv:2503.10903 [quant-ph]} \BibitemShut {NoStop}%
\bibitem [{\citenamefont {Wallraff}\ \emph {et~al.}(2004)\citenamefont {Wallraff}, \citenamefont {Schuster}, \citenamefont {Blais}, \citenamefont {Frunzio}, \citenamefont {Huang}, \citenamefont {Majer}, \citenamefont {Kumar}, \citenamefont {Girvin},\ and\ \citenamefont {Schoelkopf}}]{Wallraff2004}%
  \BibitemOpen
  \bibfield  {author} {\bibinfo {author} {\bibfnamefont {A.}~\bibnamefont {Wallraff}}, \bibinfo {author} {\bibfnamefont {D.~I.}\ \bibnamefont {Schuster}}, \bibinfo {author} {\bibfnamefont {A.}~\bibnamefont {Blais}}, \bibinfo {author} {\bibfnamefont {L.}~\bibnamefont {Frunzio}}, \bibinfo {author} {\bibfnamefont {R.-.~S.}\ \bibnamefont {Huang}}, \bibinfo {author} {\bibfnamefont {J.}~\bibnamefont {Majer}}, \bibinfo {author} {\bibfnamefont {S.}~\bibnamefont {Kumar}}, \bibinfo {author} {\bibfnamefont {S.~M.}\ \bibnamefont {Girvin}}, \ and\ \bibinfo {author} {\bibfnamefont {R.~J.}\ \bibnamefont {Schoelkopf}},\ }\href {\doibase 10.1038/nature02851} {\bibfield  {journal} {\bibinfo  {journal} {Nature}\ }\textbf {\bibinfo {volume} {431}},\ \bibinfo {pages} {162} (\bibinfo {year} {2004})}\BibitemShut {NoStop}%
\bibitem [{\citenamefont {Blais}\ \emph {et~al.}(2004)\citenamefont {Blais}, \citenamefont {Huang}, \citenamefont {Wallraff}, \citenamefont {Girvin},\ and\ \citenamefont {Schoelkopf}}]{blais2004}%
  \BibitemOpen
  \bibfield  {author} {\bibinfo {author} {\bibfnamefont {A.}~\bibnamefont {Blais}}, \bibinfo {author} {\bibfnamefont {R.-S.}\ \bibnamefont {Huang}}, \bibinfo {author} {\bibfnamefont {A.}~\bibnamefont {Wallraff}}, \bibinfo {author} {\bibfnamefont {S.~M.}\ \bibnamefont {Girvin}}, \ and\ \bibinfo {author} {\bibfnamefont {R.~J.}\ \bibnamefont {Schoelkopf}},\ }\href {\doibase 10.1103/PhysRevA.69.062320} {\bibfield  {journal} {\bibinfo  {journal} {Physical Review A}\ }\textbf {\bibinfo {volume} {69}},\ \bibinfo {pages} {062320} (\bibinfo {year} {2004})}\BibitemShut {NoStop}%
\bibitem [{\citenamefont {Song}\ \emph {et~al.}(2017)\citenamefont {Song}, \citenamefont {Xu}, \citenamefont {Liu}, \citenamefont {Yang}, \citenamefont {Zheng}, \citenamefont {Deng}, \citenamefont {Xie}, \citenamefont {Huang}, \citenamefont {Guo}, \citenamefont {Zhang} \emph {et~al.}}]{Song2017}%
  \BibitemOpen
  \bibfield  {author} {\bibinfo {author} {\bibfnamefont {C.}~\bibnamefont {Song}}, \bibinfo {author} {\bibfnamefont {K.}~\bibnamefont {Xu}}, \bibinfo {author} {\bibfnamefont {W.}~\bibnamefont {Liu}}, \bibinfo {author} {\bibfnamefont {C.-p.}\ \bibnamefont {Yang}}, \bibinfo {author} {\bibfnamefont {S.-B.}\ \bibnamefont {Zheng}}, \bibinfo {author} {\bibfnamefont {H.}~\bibnamefont {Deng}}, \bibinfo {author} {\bibfnamefont {Q.}~\bibnamefont {Xie}}, \bibinfo {author} {\bibfnamefont {K.}~\bibnamefont {Huang}}, \bibinfo {author} {\bibfnamefont {Q.}~\bibnamefont {Guo}}, \bibinfo {author} {\bibfnamefont {L.}~\bibnamefont {Zhang}},  \emph {et~al.},\ }\href {http://dx.doi.org/10.1103/PhysRevLett.119.180511} {\bibfield  {journal} {\bibinfo  {journal} {Physical Review Letters}\ }\textbf {\bibinfo {volume} {119}} (\bibinfo {year} {2017})}\BibitemShut {NoStop}%
\bibitem [{\citenamefont {Song}\ \emph {et~al.}(2019)\citenamefont {Song}, \citenamefont {Xu}, \citenamefont {Li}, \citenamefont {Zhang}, \citenamefont {Zhang}, \citenamefont {Liu}, \citenamefont {Guo}, \citenamefont {Wang}, \citenamefont {Ren}, \citenamefont {Hao} \emph {et~al.}}]{Song2019}%
  \BibitemOpen
  \bibfield  {author} {\bibinfo {author} {\bibfnamefont {C.}~\bibnamefont {Song}}, \bibinfo {author} {\bibfnamefont {K.}~\bibnamefont {Xu}}, \bibinfo {author} {\bibfnamefont {H.}~\bibnamefont {Li}}, \bibinfo {author} {\bibfnamefont {Y.-R.}\ \bibnamefont {Zhang}}, \bibinfo {author} {\bibfnamefont {X.}~\bibnamefont {Zhang}}, \bibinfo {author} {\bibfnamefont {W.}~\bibnamefont {Liu}}, \bibinfo {author} {\bibfnamefont {Q.}~\bibnamefont {Guo}}, \bibinfo {author} {\bibfnamefont {Z.}~\bibnamefont {Wang}}, \bibinfo {author} {\bibfnamefont {W.}~\bibnamefont {Ren}}, \bibinfo {author} {\bibfnamefont {J.}~\bibnamefont {Hao}},  \emph {et~al.},\ }\href {\doibase 10.1126/science.aay0600} {\bibfield  {journal} {\bibinfo  {journal} {Science}\ }\textbf {\bibinfo {volume} {365}},\ \bibinfo {pages} {574–577} (\bibinfo {year} {2019})}\BibitemShut {NoStop}%
\bibitem [{\citenamefont {Liang}\ \emph {et~al.}(soon)\citenamefont {Liang}, \citenamefont {Eberhardt},\ and\ \citenamefont {Chen}}]{yuannewarticle}%
  \BibitemOpen
  \bibfield  {author} {\bibinfo {author} {\bibfnamefont {Z.}~\bibnamefont {Liang}}, \bibinfo {author} {\bibfnamefont {J.~N.}\ \bibnamefont {Eberhardt}}, \ and\ \bibinfo {author} {\bibfnamefont {Y.-A.}\ \bibnamefont {Chen}},\ }\href@noop {} {\enquote {\bibinfo {title} {Planar quantum low-density parity-check codes with open boundaries},}\ } (\bibinfo {year} {To appear soon})\BibitemShut {NoStop}%
\bibitem [{\citenamefont {Lin}\ and\ \citenamefont {Pryadko}(2024)}]{linQuantumTwoblockGroup2023}%
  \BibitemOpen
  \bibfield  {author} {\bibinfo {author} {\bibfnamefont {H.-K.}\ \bibnamefont {Lin}}\ and\ \bibinfo {author} {\bibfnamefont {L.~P.}\ \bibnamefont {Pryadko}},\ }\href {\doibase 10.1103/PhysRevA.109.022407} {\bibfield  {journal} {\bibinfo  {journal} {Phys. Rev. A}\ }\textbf {\bibinfo {volume} {109}},\ \bibinfo {pages} {022407} (\bibinfo {year} {2024})}\BibitemShut {NoStop}%
\bibitem [{\citenamefont {Bravyi}\ \emph {et~al.}(2024)\citenamefont {Bravyi}, \citenamefont {Cross}, \citenamefont {Gambetta}, \citenamefont {Maslov}, \citenamefont {Rall},\ and\ \citenamefont {Yoder}}]{bravyi2023highthreshold}%
  \BibitemOpen
  \bibfield  {author} {\bibinfo {author} {\bibfnamefont {S.}~\bibnamefont {Bravyi}}, \bibinfo {author} {\bibfnamefont {A.~W.}\ \bibnamefont {Cross}}, \bibinfo {author} {\bibfnamefont {J.~M.}\ \bibnamefont {Gambetta}}, \bibinfo {author} {\bibfnamefont {D.}~\bibnamefont {Maslov}}, \bibinfo {author} {\bibfnamefont {P.}~\bibnamefont {Rall}}, \ and\ \bibinfo {author} {\bibfnamefont {T.~J.}\ \bibnamefont {Yoder}},\ }\href {\doibase 10.1038/s41586-024-07107-7} {\bibfield  {journal} {\bibinfo  {journal} {Nature}\ }\textbf {\bibinfo {volume} {627}},\ \bibinfo {pages} {778–782} (\bibinfo {year} {2024})}\BibitemShut {NoStop}%
\bibitem [{\citenamefont {Eberhardt}\ \emph {et~al.}(2024)\citenamefont {Eberhardt}, \citenamefont {Pereira},\ and\ \citenamefont {Steffan}}]{eberhardt2024pruningqldpccodesbivariate}%
  \BibitemOpen
  \bibfield  {author} {\bibinfo {author} {\bibfnamefont {J.~N.}\ \bibnamefont {Eberhardt}}, \bibinfo {author} {\bibfnamefont {F.~R.~F.}\ \bibnamefont {Pereira}}, \ and\ \bibinfo {author} {\bibfnamefont {V.}~\bibnamefont {Steffan}},\ }\href {https://arxiv.org/abs/2412.04181} {\enquote {\bibinfo {title} {Pruning {qLDPC} codes: Towards bivariate bicycle codes with open boundary conditions},}\ } (\bibinfo {year} {2024}),\ \Eprint {http://arxiv.org/abs/2412.04181} {arXiv:2412.04181 [quant-ph]} \BibitemShut {NoStop}%
\bibitem [{\citenamefont {Eberhardt}\ and\ \citenamefont {Steffan}(2024)}]{eberhardt2024logicaloperatorsfoldtransversalgates}%
  \BibitemOpen
  \bibfield  {author} {\bibinfo {author} {\bibfnamefont {J.~N.}\ \bibnamefont {Eberhardt}}\ and\ \bibinfo {author} {\bibfnamefont {V.}~\bibnamefont {Steffan}},\ }\href {https://arxiv.org/abs/2407.03973} {\enquote {\bibinfo {title} {Logical operators and fold-transversal gates of bivariate bicycle codes},}\ } (\bibinfo {year} {2024}),\ \Eprint {http://arxiv.org/abs/2407.03973} {arXiv:2407.03973 [quant-ph]} \BibitemShut {NoStop}%
\bibitem [{\citenamefont {Chen}\ \emph {et~al.}(2025)\citenamefont {Chen}, \citenamefont {Liu}, \citenamefont {Zhang}, \citenamefont {Liang}, \citenamefont {Chen}, \citenamefont {Liu},\ and\ \citenamefont {Song}}]{chen2025anyontheorytopologicalfrustration}%
  \BibitemOpen
  \bibfield  {author} {\bibinfo {author} {\bibfnamefont {K.}~\bibnamefont {Chen}}, \bibinfo {author} {\bibfnamefont {Y.}~\bibnamefont {Liu}}, \bibinfo {author} {\bibfnamefont {Y.}~\bibnamefont {Zhang}}, \bibinfo {author} {\bibfnamefont {Z.}~\bibnamefont {Liang}}, \bibinfo {author} {\bibfnamefont {Y.-A.}\ \bibnamefont {Chen}}, \bibinfo {author} {\bibfnamefont {K.}~\bibnamefont {Liu}}, \ and\ \bibinfo {author} {\bibfnamefont {H.}~\bibnamefont {Song}},\ }\href {https://arxiv.org/abs/2503.04699} {\enquote {\bibinfo {title} {Anyon theory and topological frustration of high-efficiency quantum {LDPC} codes},}\ } (\bibinfo {year} {2025}),\ \Eprint {http://arxiv.org/abs/2503.04699} {arXiv:2503.04699 [quant-ph]} \BibitemShut {NoStop}%
\bibitem [{\citenamefont {Liang}\ \emph {et~al.}(2025)\citenamefont {Liang}, \citenamefont {Liu}, \citenamefont {Song},\ and\ \citenamefont {Chen}}]{liang2025generalizedtoriccodestwisted}%
  \BibitemOpen
  \bibfield  {author} {\bibinfo {author} {\bibfnamefont {Z.}~\bibnamefont {Liang}}, \bibinfo {author} {\bibfnamefont {K.}~\bibnamefont {Liu}}, \bibinfo {author} {\bibfnamefont {H.}~\bibnamefont {Song}}, \ and\ \bibinfo {author} {\bibfnamefont {Y.-A.}\ \bibnamefont {Chen}},\ }\href {https://arxiv.org/abs/2503.03827} {\enquote {\bibinfo {title} {Generalized toric codes on twisted tori for quantum error correction},}\ } (\bibinfo {year} {2025}),\ \Eprint {http://arxiv.org/abs/2503.03827} {arXiv:2503.03827 [quant-ph]} \BibitemShut {NoStop}%
\bibitem [{\citenamefont {Liang}\ \emph {et~al.}(2024)\citenamefont {Liang}, \citenamefont {Yang}, \citenamefont {Iosue},\ and\ \citenamefont {Chen}}]{liang2024operatoralgebraalgorithmicconstruction}%
  \BibitemOpen
  \bibfield  {author} {\bibinfo {author} {\bibfnamefont {Z.}~\bibnamefont {Liang}}, \bibinfo {author} {\bibfnamefont {B.}~\bibnamefont {Yang}}, \bibinfo {author} {\bibfnamefont {J.~T.}\ \bibnamefont {Iosue}}, \ and\ \bibinfo {author} {\bibfnamefont {Y.-A.}\ \bibnamefont {Chen}},\ }\href {https://arxiv.org/abs/2410.11942} {\enquote {\bibinfo {title} {Operator algebra and algorithmic construction of boundaries and defects in (2+1){D} topological pauli stabilizer codes},}\ } (\bibinfo {year} {2024}),\ \Eprint {http://arxiv.org/abs/2410.11942} {arXiv:2410.11942 [quant-ph]} \BibitemShut {NoStop}%
\bibitem [{\citenamefont {Berthusen}\ \emph {et~al.}(2025)\citenamefont {Berthusen}, \citenamefont {Devulapalli}, \citenamefont {Schoute}, \citenamefont {Childs}, \citenamefont {Gullans}, \citenamefont {Gorshkov},\ and\ \citenamefont {Gottesman}}]{Berthusen_2025}%
  \BibitemOpen
  \bibfield  {author} {\bibinfo {author} {\bibfnamefont {N.}~\bibnamefont {Berthusen}}, \bibinfo {author} {\bibfnamefont {D.}~\bibnamefont {Devulapalli}}, \bibinfo {author} {\bibfnamefont {E.}~\bibnamefont {Schoute}}, \bibinfo {author} {\bibfnamefont {A.~M.}\ \bibnamefont {Childs}}, \bibinfo {author} {\bibfnamefont {M.~J.}\ \bibnamefont {Gullans}}, \bibinfo {author} {\bibfnamefont {A.~V.}\ \bibnamefont {Gorshkov}}, \ and\ \bibinfo {author} {\bibfnamefont {D.}~\bibnamefont {Gottesman}},\ }\href {\doibase 10.1103/prxquantum.6.010306} {\bibfield  {journal} {\bibinfo  {journal} {PRX Quantum}\ }\textbf {\bibinfo {volume} {6}} (\bibinfo {year} {2025}),\ 10.1103/prxquantum.6.010306}\BibitemShut {NoStop}%
\bibitem [{\citenamefont {Pattison}\ \emph {et~al.}(2023)\citenamefont {Pattison}, \citenamefont {Krishna},\ and\ \citenamefont {Preskill}}]{pattison2023hierarchicalmemoriessimulatingquantum}%
  \BibitemOpen
  \bibfield  {author} {\bibinfo {author} {\bibfnamefont {C.~A.}\ \bibnamefont {Pattison}}, \bibinfo {author} {\bibfnamefont {A.}~\bibnamefont {Krishna}}, \ and\ \bibinfo {author} {\bibfnamefont {J.}~\bibnamefont {Preskill}},\ }\href {https://arxiv.org/abs/2303.04798} {\enquote {\bibinfo {title} {Hierarchical memories: Simulating quantum ldpc codes with local gates},}\ } (\bibinfo {year} {2023}),\ \Eprint {http://arxiv.org/abs/2303.04798} {arXiv:2303.04798 [quant-ph]} \BibitemShut {NoStop}%
\bibitem [{\citenamefont {Haah}(2017)}]{Haah_2017}%
  \BibitemOpen
  \bibfield  {author} {\bibinfo {author} {\bibfnamefont {J.}~\bibnamefont {Haah}},\ }\href {\doibase 10.15446/recolma.v50n2.62214} {\bibfield  {journal} {\bibinfo  {journal} {Revista Colombiana de Matemáticas}\ }\textbf {\bibinfo {volume} {50}},\ \bibinfo {pages} {299} (\bibinfo {year} {2017})}\BibitemShut {NoStop}%
\bibitem [{\citenamefont {Pecorari}\ \emph {et~al.}(2024)\citenamefont {Pecorari}, \citenamefont {Jandura}, \citenamefont {Brennen},\ and\ \citenamefont {Pupillo}}]{pecorari2024highrate}%
  \BibitemOpen
  \bibfield  {author} {\bibinfo {author} {\bibfnamefont {L.}~\bibnamefont {Pecorari}}, \bibinfo {author} {\bibfnamefont {S.}~\bibnamefont {Jandura}}, \bibinfo {author} {\bibfnamefont {G.~K.}\ \bibnamefont {Brennen}}, \ and\ \bibinfo {author} {\bibfnamefont {G.}~\bibnamefont {Pupillo}},\ }\href@noop {} {\enquote {\bibinfo {title} {High-rate quantum {LDPC} codes for long-range-connected neutral atom registers},}\ } (\bibinfo {year} {2024}),\ \Eprint {http://arxiv.org/abs/2404.13010} {arXiv:2404.13010 [quant-ph]} \BibitemShut {NoStop}%
\bibitem [{GAP()}]{GAP4}%
  \BibitemOpen
  GAP,\ \href@noop {} {\emph {\bibinfo {title} {{GAP -- Groups, Algorithms, and Programming, Version 4.13.1}}}},\ \bibinfo {organization} {The GAP~Group} (\bibinfo {year} {2024})\BibitemShut {NoStop}%
\bibitem [{\citenamefont {Pryadko}\ \emph {et~al.}(2022)\citenamefont {Pryadko}, \citenamefont {Shabashov},\ and\ \citenamefont {Kozin}}]{Pryadko2022}%
  \BibitemOpen
  \bibfield  {author} {\bibinfo {author} {\bibfnamefont {L.~P.}\ \bibnamefont {Pryadko}}, \bibinfo {author} {\bibfnamefont {V.~A.}\ \bibnamefont {Shabashov}}, \ and\ \bibinfo {author} {\bibfnamefont {V.~K.}\ \bibnamefont {Kozin}},\ }\href {\doibase 10.21105/joss.04120} {\bibfield  {journal} {\bibinfo  {journal} {Journal of Open Source Software}\ }\textbf {\bibinfo {volume} {7}},\ \bibinfo {pages} {4120} (\bibinfo {year} {2022})}\BibitemShut {NoStop}%
\end{thebibliography}%

\begin{widetext}
\newpage
    \appendix

\section{The hypergraph product case}\label{sec: appendix hgp}

The tile codes we constructed can be viewed as planar versions of BB codes. First attempts to find planar versions of BB codes -- and with that, tile codes -- have been achieved in~\cite{eberhardt2024pruningqldpccodesbivariate, pecorari2024highrate}. We refer for details on BB codes to~\cite{linQuantumTwoblockGroup2023,eberhardt2024pruningqldpccodesbivariate}. We mention that the tile codes constructed in Refs.~\cite{eberhardt2024pruningqldpccodesbivariate, pecorari2024highrate} are exactly the planar version of BB codes that are hypergraph product codes. 
We will now explain how all of the tile codes constructed in~\cite{eberhardt2024pruningqldpccodesbivariate, pecorari2024highrate} naturally arise by our simple construction. 

The stabilizer tiles are specified by a pair of polynomials $a(x)$ and $b(y)$. The stabilizer generators will fit into a box of size $(\text{deg}(a) + 1) \times (\text{deg} (b) + 1)$. For the $\op X$-type stabilizer tile, the polynomial $a(x)$ describes which of the horizontal edges in the highest layer of the box are colored, and the polynomial $b(y)$ which of the rightmost vertical edges. For the $\op Z$-type stabilizer, $a(x)$ describes which of the vertical edges on the lowest layer are colored and which of the horizontal edges on the leftmost layer. For example, if $a(x) = 1 + x + x^2$ and $b(y) = 1 + y + y^2$, the stabilizer tiles look as follows.

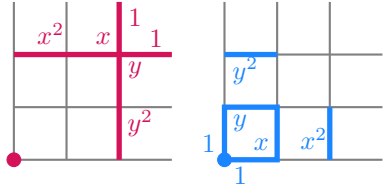
\begin{equation*}
    \begin{tikzpicture}[scale=0.7]
        \begin{scope}[xshift=4cm]
            \draw[\latticecolor, \latticethickness] (3,0) -- (0,0) -- (0,3) ;
            \draw[\latticecolor, \latticethickness] (1,0) -- (1,3);
            \draw[\latticecolor, \latticethickness] (2,0) -- (2,3);
            \draw[\latticecolor, \latticethickness] (0,1) -- (3,1);
            \draw[\latticecolor, \latticethickness] (0,2) -- (3,2);
            
            \draw[good_blue, line width = 2pt] (0,0) -- (0,1) --(1,1) -- (1,0) -- cycle;
            \draw[good_blue, line width = 2pt] (0,2) -- (1,2);
            \draw[good_blue, line width = 2pt] (2,0) -- (2,1);
            \draw[fill=good_blue, color = good_blue] (0,0) circle (0.13);

            \node[color = good_blue] at (-.3,.3) {$1$};
            \node[color = good_blue] at (.7,.3) {$x$};
            \node[color = good_blue] at (1.7,.4) {$x^2$};

            \node[color = good_blue] at (.3,-.3) {$1$};
            \node[color = good_blue] at (.3,.7) {$y$};
            \node[color = good_blue] at (.4,1.7) {$y^2$};
        \end{scope}
        
        \begin{scope}[xshift=0cm]
            \draw[\latticecolor, \latticethickness] (3,0) -- (0,0) -- (0,3) ;
            \draw[\latticecolor, \latticethickness] (1,0) -- (1,3);
            \draw[\latticecolor, \latticethickness] (2,0) -- (2,3);
            \draw[\latticecolor, \latticethickness] (0,1) -- (3,1);
            \draw[\latticecolor, \latticethickness] (0,2) -- (3,2);
            
            \draw[good_red, line width = 2pt] (2,0) -- (2,3);
            \draw[good_red, line width = 2pt] (0,2) -- (3,2);
            
            \draw[fill=good_red, color=good_red] (0,0) circle (0.13);

            \node[color = good_red] at (2.7,2.3) {$1$};
            \node[color = good_red] at (1.7,2.3) {$x$};
            \node[color = good_red] at (.7,2.4) {$x^2$};

            \node[color = good_red] at (2.3,2.7) {$1$};
            \node[color = good_red] at (2.3,1.7) {$y$};
            \node[color = good_red] at (2.4,.7) {$y^2$};
        \end{scope}
    \end{tikzpicture}
\end{equation*}

BB codes arise by tiling a torus with these stabilizer tiles. In~\cite{eberhardt2024pruningqldpccodesbivariate, pecorari2024highrate}, the authors try to find planar versions of these codes. We will now demonstrate that all codes they find can be recovered using our simple construction. For a visualization of the construction, see \Cref{fig:visualization of algorithm}.

While it is straightforward to generalize this construction, we will for convenience assume $a$ and $b$ having the same monomial terms. In this case, the minimal possible size of a box supporting the $\op X$- and $\op Z$-type stabilizers will be $B = \text{deg}(a) + 1$. 

\begin{figure*}[h!]
    \centering
    \begin{tikzpicture}[scale=0.7]
        \begin{scope}[xshift = 1cm]
            \foreach \x in {0,...,5} {
            \draw[white, thin] (\x/2,0) -- (\x/2,3);
        }
        
        \foreach \y in {0,...,5} {
            \draw[white, thin] (0,\y/2) -- (3,\y/2);
            \foreach \x in {0,...,3} {
                \foreach \y in {0,...,3} {
                    \draw[fill=black] (\x/2,\y/2) circle (0.1);
                }
            }
        }
        \end{scope}

        \begin{scope}[xshift=6cm]
            \foreach \x in {0,...,5} {
                \draw[gray, thin] (\x/2,0) -- (\x/2,3);
            }
            
            \foreach \y in {0,...,5} {
                \draw[gray, thin] (0,\y/2) -- (3,\y/2);
            }
            
            \foreach \x in {0,...,3} {
                \foreach \y in {0,...,3} {
                    \draw[fill=black] (\x/2,\y/2) circle (0.1);
                }
            }
            \foreach \x in {0,...,3} {
                \foreach \y in {0,...,3} {
                    \draw[fill=black] (\x/2,\y/2) circle (0.1);
                }
            }
        \end{scope}

        \begin{scope}[xshift=12cm]
            \foreach \x in {0,...,5} {
                \draw[gray, thin] (\x/2,0) -- (\x/2,3);
            }
            
            \foreach \y in {0,...,5} {
                \draw[gray, thin] (0,\y/2) -- (3,\y/2);
            }
            
            \foreach \x in {0,...,3} {
                \foreach \y in {0,...,3} {
                    \draw[fill=black] (\x/2,\y/2) circle (0.1);
                }
            }
            
            \foreach \x in {0,...,3} {
                \draw[fill=good_red, color = good_red] (\x/2,-0.5) circle (0.1);
                \draw[fill=good_red, color = good_red] (\x/2,-1) circle (0.1);
                \draw[fill=good_red, color = good_red] (\x/2,2) circle (0.1);
                \draw[fill=good_red, color = good_red] (\x/2,2.5) circle (0.1);
            }
            
            \foreach \y in {0,...,3} {
                \draw[fill=good_blue,color = good_blue] (-0.5,\y/2) circle (0.1);
                \draw[fill=good_blue, color = good_blue] (-1,\y/2) circle (0.1);
                \draw[fill=good_blue, color = good_blue] (2,\y/2) circle (0.1);
                \draw[fill=good_blue, color = good_blue] (2.5,\y/2) circle (0.1);
            }
        \end{scope}

        \begin{scope}[xshift=18cm]
            \foreach \x in {2,...,5} {
                \draw[gray, thin] (\x/2,0) -- (\x/2,1.5);
            }
            
            \foreach \y in {0,...,5} {
                \draw[gray, thin] (0,\y/2) -- (3,\y/2);
            }
            
            \foreach \x in {0,...,3} {
                \foreach \y in {0,...,3} {
                    \draw[fill=black] (\x/2,\y/2) circle (0.1);
                }
            }
            
            \foreach \x in {0,...,3} {
                \draw[fill=good_red, color = good_red] (\x/2,-0.5) circle (0.1);
                \draw[fill=good_red, color = good_red] (\x/2,-1) circle (0.1);
            }
            
            \foreach \y in {0,...,3} {
                \draw[fill=good_blue, color = good_blue] (2,\y/2) circle (0.1);
                \draw[fill=good_blue, color = good_blue] (2.5,\y/2) circle (0.1);
            }
        \end{scope}

        \begin{scope}[yshift=-2cm]
            \draw[->, thick] (0,0) -- (20,0);
            \foreach \x/\t in {2.5/1, 8/2, 14/3, 19.5/4} {
                \draw[thick] (\x,-0.2) -- (\x,0.2);
                \node[below] at (\x,-0.2) {Step \t};
            }
            
            \node[below, text width=4cm, align=center] at (2.5,-1) {Pick layout of \\ bulk stabilizers};
            \node[below, text width=4cm, align=center] at (8,-1) {Draw qubit layout \\ accordingly};
            \node[below, text width=4cm, align=center] at (14,-1) {Add boundary \\stabilizers};
            \node[below, text width=4cm, align=center] at (19.5,-1) {Remove unnecessary \\ qubits and stabilizers};
        \end{scope}
    \end{tikzpicture}
    \caption{Visualization of the construction in the hypergraph product set up. Following the simple steps of choosing a rectangular layout of bulk stabilizers and following the construction in the main text recovers all constructions from~\cite{eberhardt2024pruningqldpccodesbivariate, pecorari2024highrate}.}
    \label{fig:visualization of algorithm}
\end{figure*}
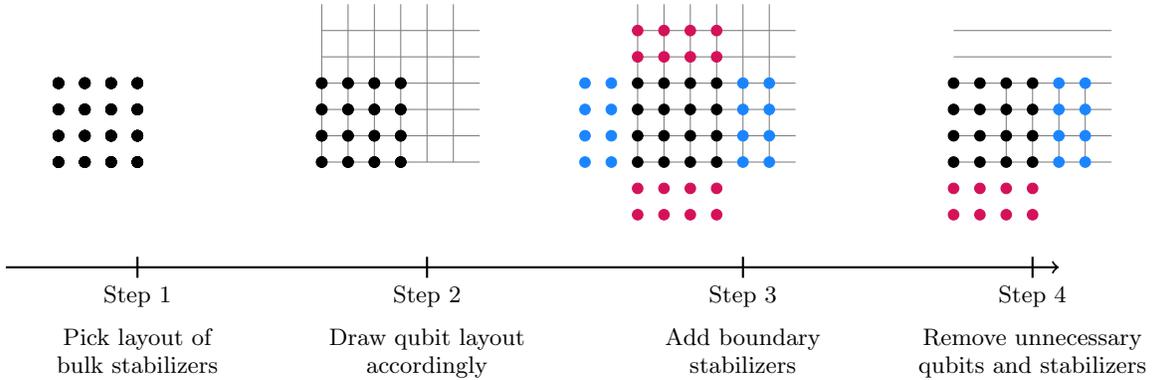

In the first step of our construction, we draw a $(l-(B-1))\times (m-(B-1))$ lattice of bulk stabilizers which in the next step will lead -- because of our choice of stabilizer tiles -- to a $l \times m$ lattice of qubits. Then, we place $(B-1)$ layers of $\op X$-type stabilizer on the bottom and top and $(B-1)$ layers of $\op Z$-type stabilizers on the left and right.
From the shape of the $\op X$- and $\op Z$-type stabilizers it is clear that the uppermost $B-1$ layers of vertical qubits are not supported by any $\op Z$-type stabilizer and the leftmost $B-1$ layers of vertical qubits are not contained in the support of any $\op X$-type stabilizer. In Step 4 of the algorithm, we therefore remove these qubits as also depicted in~\ref{fig:visualization of algorithm}. Again, carefully examining the shape of the stabilizers we observe that the uppermost $B-1$ layers of $\op X$-type stabilizers as well as the leftmost $B-1$ layers of $\op Z$-type stabilizers have empty support since all of their support has just been removed. Consequently, these stabilizers will be removed as well in Step 4. It is clear that all remaining stabilizers have non-empty support and are independent. The final code has $2lm - l(B-1) - m(B-1) = lm + (l - \deg(a))(m - \deg(b))$ physical qubits and $(l -(B-1))(m - (B - 1)) + (l - (B - 1))(B-1) + (m - (B - 1)) (B - 1)$ independent stabilizer generators, hence $(B-1)^2 = \text{deg}(a)^2$ logical qubits. It is clear that starting with two different polynomials $a$ and $b$ and using rectangular boxes instead of square boxes, this analysis demonstrates a logical dimension of $\text{deg}(a) \text{deg}(b)$ which is precisely the result of~\cite{eberhardt2024pruningqldpccodesbivariate, pecorari2024highrate}. It is also clear that our algorithm removes whole rows or columns of qubits and stabilizers, that is, the resulting code is still a hypergraph product code.

\section{Stabilizers of weight 6 confined in $3\times 3$ boxes yielding $[[288, 8, 12]]$ codes}\label{sec:appendix weight 6}

We exhaustively searched over all possible weight-6 stabilizers confined in a $3 \times 3$ tile. We find that there are 16 different codes with these parameters.
All of those stabilizers yielding $[[288, 8,12]]$ codes are depicted in \Cref{tab:w6_all}.

\begin{table}[h]
    \centering
    \begin{tabular}{c|c|c|c}
            \begin{tikzpicture}[scale=0.5]
        \def\horizontalqubits{(0,0),(0,1), (2,2)}
        \def\verticalqubits{(0,2),(1,0),(2,0)}

        \def\horizontalqubitsz{
            (\boxsize - 1 - 0,\boxsize - 1 - 2),
            (\boxsize - 1 - 1,\boxsize - 1 - 0),
            (\boxsize - 1 - 2,\boxsize - 1 - 0)
        }
        \def\verticalqubitsz{
            (\boxsize - 1 - 0,\boxsize - 1 - 0),
            (\boxsize - 1 - 0,\boxsize - 1 - 1),
            (\boxsize - 1 - 2,\boxsize - 1 - 2)
        }
        \def\boxsize{3}
        
        \begin{scope}[xshift=0cm]
            \draw[\latticecolor, \latticethickness] (3,0) -- (0,0) -- (0,3) ;
            \draw[\latticecolor, \latticethickness] (1,0) -- (1,3);
            \draw[\latticecolor, \latticethickness] (2,0) -- (2,3);
            \draw[\latticecolor, \latticethickness] (0,1) -- (3,1);
            \draw[\latticecolor, \latticethickness] (0,2) -- (3,2);
            \fill[color = good_red, opacity = .0] (0,0) -- (3,0) -- (3,3) -- (0,3) -- cycle;
            \foreach \pos in \horizontalqubits {
                \draw[good_red, line width=2pt] \pos -- ++(1,0);
            }
            \foreach \pos in \verticalqubits {
                \draw[good_red, line width=2pt] \pos -- ++(0,1);
            }
            
            \draw[fill=good_red, color = good_red] (0,0) circle (0.15);
        \end{scope}

        \def\horizontalqubits{(0,0),(0,1), (2,2)}
        \def\verticalqubits{(0,2),(1,0),(2,0)}
        \begin{scope}[xshift=4cm]
            \draw[\latticecolor, \latticethickness] (3,0) -- (0,0) -- (0,3) ;
            \draw[\latticecolor, \latticethickness] (1,0) -- (1,3);
            \draw[\latticecolor, \latticethickness] (2,0) -- (2,3);
            \draw[\latticecolor, \latticethickness] (0,1) -- (3,1);
            \draw[\latticecolor, \latticethickness] (0,2) -- (3,2);
            
            \foreach \pos in \horizontalqubitsz {
                \draw[good_blue, line width=2pt] \pos -- ++(1,0);
            }
            \foreach \pos in \verticalqubitsz {
                \draw[good_blue, line width=2pt] \pos -- ++(0,1);
            }
            
            \draw[fill=good_blue, color = good_blue] (0,0) circle (0.15);
        \end{scope}
        \addvmargin{1mm}
    \end{tikzpicture} \;\;\;&\;\;\;   
    \begin{tikzpicture}[scale=0.5]
        \def\horizontalqubits{(0,0),(0,1), (2,2)}
        \def\verticalqubits{(0,2),(1,2),(2,0)}

        \def\horizontalqubitsz{
        (\boxsize - 1 - 0,\boxsize - 1 - 2),
        (\boxsize - 1 - 1,\boxsize - 1 - 2),
        (\boxsize - 1 - 2,\boxsize - 1 - 0)}
        \def\verticalqubitsz{
        (\boxsize - 1 - 0,\boxsize - 1 - 0),
        (\boxsize - 1 - 0,\boxsize - 1 - 1),
        (\boxsize - 1 - 2,\boxsize - 1 - 2)}
        \def\boxsize{3}
        
        \begin{scope}[xshift=0cm]
            \draw[\latticecolor, \latticethickness] (3,0) -- (0,0) -- (0,3) ;
            \draw[\latticecolor, \latticethickness] (1,0) -- (1,3);
            \draw[\latticecolor, \latticethickness] (2,0) -- (2,3);
            \draw[\latticecolor, \latticethickness] (0,1) -- (3,1);
            \draw[\latticecolor, \latticethickness] (0,2) -- (3,2);
            \fill[color = good_red, opacity = .0] (0,0) -- (3,0) -- (3,3) -- (0,3) -- cycle;
            \foreach \pos in \horizontalqubits {
                \draw[good_red, line width=2pt] \pos -- ++(1,0);
            }
            \foreach \pos in \verticalqubits {
                \draw[good_red, line width=2pt] \pos -- ++(0,1);
            }
            
            \draw[fill=good_red, color = good_red] (0,0) circle (0.15);
        \end{scope}

        \def\horizontalqubits{(0,0),(0,1), (2,2)}
        \def\verticalqubits{(0,2),(1,0),(2,0)}
        \begin{scope}[xshift=4cm]
            \draw[\latticecolor, \latticethickness] (3,0) -- (0,0) -- (0,3) ;
            \draw[\latticecolor, \latticethickness] (1,0) -- (1,3);
            \draw[\latticecolor, \latticethickness] (2,0) -- (2,3);
            \draw[\latticecolor, \latticethickness] (0,1) -- (3,1);
            \draw[\latticecolor, \latticethickness] (0,2) -- (3,2);
            
            \foreach \pos in \horizontalqubitsz {
                \draw[good_blue, line width=2pt] \pos -- ++(1,0);
            }
            \foreach \pos in \verticalqubitsz {
                \draw[good_blue, line width=2pt] \pos -- ++(0,1);
            }
            
            \draw[fill=good_blue, color = good_blue] (0,0) circle (0.15);
        \end{scope}
    \end{tikzpicture} \;\;\;& \;\;\;             
    \begin{tikzpicture}[scale=0.5]
        \def\horizontalqubits{(0,0),(1,0), (2,2)}
        \def\verticalqubits{(0,1),(0,2),(2,0)}

        \def\horizontalqubitsz{
        (\boxsize - 1 - 0,\boxsize - 1 - 1),
        (\boxsize - 1 - 0,\boxsize - 1 - 2),
        (\boxsize - 1 - 2,\boxsize - 1 - 0)
        }
        \def\verticalqubitsz{
        (\boxsize - 1 - 0,\boxsize - 1 - 0),
        (\boxsize - 1 - 1,\boxsize - 1 - 0),
        (\boxsize - 1 - 2,\boxsize - 1 - 2)
        }
        \def\boxsize{3}
        
        \begin{scope}[xshift=0cm]
            \draw[\latticecolor, \latticethickness] (3,0) -- (0,0) -- (0,3) ;
            \draw[\latticecolor, \latticethickness] (1,0) -- (1,3);
            \draw[\latticecolor, \latticethickness] (2,0) -- (2,3);
            \draw[\latticecolor, \latticethickness] (0,1) -- (3,1);
            \draw[\latticecolor, \latticethickness] (0,2) -- (3,2);
            \fill[color = good_red, opacity = .0] (0,0) -- (3,0) -- (3,3) -- (0,3) -- cycle;
            \foreach \pos in \horizontalqubits {
                \draw[good_red, line width=2pt] \pos -- ++(1,0);
            }
            \foreach \pos in \verticalqubits {
                \draw[good_red, line width=2pt] \pos -- ++(0,1);
            }
            
            \draw[fill=good_red, color = good_red] (0,0) circle (0.15);
        \end{scope}

        \def\horizontalqubits{(0,0),(0,1), (2,2)}
        \def\verticalqubits{(0,2),(1,0),(2,0)}
        \begin{scope}[xshift=4cm]
            \draw[\latticecolor, \latticethickness] (3,0) -- (0,0) -- (0,3) ;
            \draw[\latticecolor, \latticethickness] (1,0) -- (1,3);
            \draw[\latticecolor, \latticethickness] (2,0) -- (2,3);
            \draw[\latticecolor, \latticethickness] (0,1) -- (3,1);
            \draw[\latticecolor, \latticethickness] (0,2) -- (3,2);
            
            \foreach \pos in \horizontalqubitsz {
                \draw[good_blue, line width=2pt] \pos -- ++(1,0);
            }
            \foreach \pos in \verticalqubitsz {
                \draw[good_blue, line width=2pt] \pos -- ++(0,1);
            }
            
            \draw[fill=good_blue, color = good_blue] (0,0) circle (0.15);
        \end{scope}
    \end{tikzpicture}\;\;\;&   \;\;\;          
    \begin{tikzpicture}[scale=0.5]
        \def\horizontalqubits{(0,0),(1,0), (2,2)}
        \def\verticalqubits{(0,2),(2,0),(2,1)}

        \def\horizontalqubitsz{
        (\boxsize - 1 - 0,\boxsize - 1 - 2),
        (\boxsize - 1 - 2,\boxsize - 1 - 0),
        (\boxsize - 1 - 2,\boxsize - 1 - 1)}
        \def\verticalqubitsz{
        (\boxsize - 1 - 0,\boxsize - 1 - 0),
        (\boxsize - 1 - 1,\boxsize - 1 - 0),
        (\boxsize - 1 - 2,\boxsize - 1 - 2)}
        \def\boxsize{3}
        
        \begin{scope}[xshift=0cm]
            \draw[\latticecolor, \latticethickness] (3,0) -- (0,0) -- (0,3) ;
            \draw[\latticecolor, \latticethickness] (1,0) -- (1,3);
            \draw[\latticecolor, \latticethickness] (2,0) -- (2,3);
            \draw[\latticecolor, \latticethickness] (0,1) -- (3,1);
            \draw[\latticecolor, \latticethickness] (0,2) -- (3,2);
            \fill[color = good_red, opacity = .0] (0,0) -- (3,0) -- (3,3) -- (0,3) -- cycle;
            \foreach \pos in \horizontalqubits {
                \draw[good_red, line width=2pt] \pos -- ++(1,0);
            }
            \foreach \pos in \verticalqubits {
                \draw[good_red, line width=2pt] \pos -- ++(0,1);
            }
            
            \draw[fill=good_red, color = good_red] (0,0) circle (0.15);
        \end{scope}

        \def\horizontalqubits{(0,0),(0,1), (2,2)}
        \def\verticalqubits{(0,2),(1,0),(2,0)}
        \begin{scope}[xshift=4cm]
            \draw[\latticecolor, \latticethickness] (3,0) -- (0,0) -- (0,3) ;
            \draw[\latticecolor, \latticethickness] (1,0) -- (1,3);
            \draw[\latticecolor, \latticethickness] (2,0) -- (2,3);
            \draw[\latticecolor, \latticethickness] (0,1) -- (3,1);
            \draw[\latticecolor, \latticethickness] (0,2) -- (3,2);
            
            \foreach \pos in \horizontalqubitsz {
                \draw[good_blue, line width=2pt] \pos -- ++(1,0);
            }
            \foreach \pos in \verticalqubitsz {
                \draw[good_blue, line width=2pt] \pos -- ++(0,1);
            }
            
            \draw[fill=good_blue, color = good_blue] (0,0) circle (0.15);
        \end{scope}
    \end{tikzpicture} \\
    \hline 
    \begin{tikzpicture}[scale=0.5]
        \def\horizontalqubits{(0,0),(1,2), (2,2)}
        \def\verticalqubits{(0,1),(0,2),(2,0)}

        \def\horizontalqubitsz{
        (\boxsize - 1 - 0,\boxsize - 1 - 1),
        (\boxsize - 1 - 0,\boxsize - 1 - 2),
        (\boxsize - 1 - 2,\boxsize - 1 - 0)}
        \def\verticalqubitsz{
        (\boxsize - 1 - 0,\boxsize - 1 - 0),
        (\boxsize - 1 - 1,\boxsize - 1 - 2),
        (\boxsize - 1 - 2,\boxsize - 1 - 2)}
        \def\boxsize{3}
        
        \begin{scope}[xshift=0cm]
            \draw[\latticecolor, \latticethickness] (3,0) -- (0,0) -- (0,3) ;
            \draw[\latticecolor, \latticethickness] (1,0) -- (1,3);
            \draw[\latticecolor, \latticethickness] (2,0) -- (2,3);
            \draw[\latticecolor, \latticethickness] (0,1) -- (3,1);
            \draw[\latticecolor, \latticethickness] (0,2) -- (3,2);
            \fill[color = good_red, opacity = .0] (0,0) -- (3,0) -- (3,3) -- (0,3) -- cycle;
            \foreach \pos in \horizontalqubits {
                \draw[good_red, line width=2pt] \pos -- ++(1,0);
            }
            \foreach \pos in \verticalqubits {
                \draw[good_red, line width=2pt] \pos -- ++(0,1);
            }
            
            \draw[fill=good_red, color = good_red] (0,0) circle (0.15);
        \end{scope}

        \def\horizontalqubits{(0,0),(0,1), (2,2)}
        \def\verticalqubits{(0,2),(1,0),(2,0)}
        \begin{scope}[xshift=4cm]
            \draw[\latticecolor, \latticethickness] (3,0) -- (0,0) -- (0,3) ;
            \draw[\latticecolor, \latticethickness] (1,0) -- (1,3);
            \draw[\latticecolor, \latticethickness] (2,0) -- (2,3);
            \draw[\latticecolor, \latticethickness] (0,1) -- (3,1);
            \draw[\latticecolor, \latticethickness] (0,2) -- (3,2);
            
            \foreach \pos in \horizontalqubitsz {
                \draw[good_blue, line width=2pt] \pos -- ++(1,0);
            }
            \foreach \pos in \verticalqubitsz {
                \draw[good_blue, line width=2pt] \pos -- ++(0,1);
            }
            
            \draw[fill=good_blue, color = good_blue] (0,0) circle (0.15);
        \end{scope}
    \end{tikzpicture} \;\;\;&  \;\;\;           
    \begin{tikzpicture}[scale=0.5]
        \def\horizontalqubits{(0,0),(1,2), (2,2)}
        \def\verticalqubits{(0,2),(2,0),(2,1)}

        \def\horizontalqubitsz{
        (\boxsize - 1 - 0,\boxsize - 1 - 2),
        (\boxsize - 1 - 2,\boxsize - 1 - 0),
        (\boxsize - 1 - 2,\boxsize - 1 - 1)
        }
        \def\verticalqubitsz{
        (\boxsize - 1 - 0,\boxsize - 1 - 0),
        (\boxsize - 1 - 1,\boxsize - 1 - 2),
        (\boxsize - 1 - 2,\boxsize - 1 - 2)}
        \def\boxsize{3}
        
        \begin{scope}[xshift=0cm]
            \draw[\latticecolor, \latticethickness] (3,0) -- (0,0) -- (0,3) ;
            \draw[\latticecolor, \latticethickness] (1,0) -- (1,3);
            \draw[\latticecolor, \latticethickness] (2,0) -- (2,3);
            \draw[\latticecolor, \latticethickness] (0,1) -- (3,1);
            \draw[\latticecolor, \latticethickness] (0,2) -- (3,2);
            \fill[color = good_red, opacity = .0] (0,0) -- (3,0) -- (3,3) -- (0,3) -- cycle;
            \foreach \pos in \horizontalqubits {
                \draw[good_red, line width=2pt] \pos -- ++(1,0);
            }
            \foreach \pos in \verticalqubits {
                \draw[good_red, line width=2pt] \pos -- ++(0,1);
            }
            
            \draw[fill=good_red, color = good_red] (0,0) circle (0.15);
        \end{scope}

        \def\horizontalqubits{(0,0),(0,1), (2,2)}
        \def\verticalqubits{(0,2),(1,0),(2,0)}
        \begin{scope}[xshift=4cm]
            \draw[\latticecolor, \latticethickness] (3,0) -- (0,0) -- (0,3) ;
            \draw[\latticecolor, \latticethickness] (1,0) -- (1,3);
            \draw[\latticecolor, \latticethickness] (2,0) -- (2,3);
            \draw[\latticecolor, \latticethickness] (0,1) -- (3,1);
            \draw[\latticecolor, \latticethickness] (0,2) -- (3,2);
            
            \foreach \pos in \horizontalqubitsz {
                \draw[good_blue, line width=2pt] \pos -- ++(1,0);
            }
            \foreach \pos in \verticalqubitsz {
                \draw[good_blue, line width=2pt] \pos -- ++(0,1);
            }
            
            \draw[fill=good_blue, color = good_blue] (0,0) circle (0.15);
        \end{scope}
    \end{tikzpicture} \;\;\;&  \;\;\;            
    \begin{tikzpicture}[scale=0.5]
        \def\horizontalqubits{(0,0),(2,1), (2,2)}
        \def\verticalqubits{(0,2),(1,0),(2,0)}

        \def\horizontalqubitsz{
        (\boxsize - 1 - 0,\boxsize - 1 - 2),
        (\boxsize - 1 - 1,\boxsize - 1 - 0),
        (\boxsize - 1 - 2,\boxsize - 1 - 0)}
        \def\verticalqubitsz{
        (\boxsize - 1 - 0,\boxsize - 1 - 0),
        (\boxsize - 1 - 2,\boxsize - 1 - 1),
        (\boxsize - 1 - 2,\boxsize - 1 - 2)}
        \def\boxsize{3}
        
        \begin{scope}[xshift=0cm]
            \draw[\latticecolor, \latticethickness] (3,0) -- (0,0) -- (0,3) ;
            \draw[\latticecolor, \latticethickness] (1,0) -- (1,3);
            \draw[\latticecolor, \latticethickness] (2,0) -- (2,3);
            \draw[\latticecolor, \latticethickness] (0,1) -- (3,1);
            \draw[\latticecolor, \latticethickness] (0,2) -- (3,2);
            \fill[color = good_red, opacity = .0] (0,0) -- (3,0) -- (3,3) -- (0,3) -- cycle;
            \foreach \pos in \horizontalqubits {
                \draw[good_red, line width=2pt] \pos -- ++(1,0);
            }
            \foreach \pos in \verticalqubits {
                \draw[good_red, line width=2pt] \pos -- ++(0,1);
            }
            
            \draw[fill=good_red, color = good_red] (0,0) circle (0.15);
        \end{scope}

        \def\horizontalqubits{(0,0),(0,1), (2,2)}
        \def\verticalqubits{(0,2),(1,0),(2,0)}
        \begin{scope}[xshift=4cm]
            \draw[\latticecolor, \latticethickness] (3,0) -- (0,0) -- (0,3) ;
            \draw[\latticecolor, \latticethickness] (1,0) -- (1,3);
            \draw[\latticecolor, \latticethickness] (2,0) -- (2,3);
            \draw[\latticecolor, \latticethickness] (0,1) -- (3,1);
            \draw[\latticecolor, \latticethickness] (0,2) -- (3,2);
            
            \foreach \pos in \horizontalqubitsz {
                \draw[good_blue, line width=2pt] \pos -- ++(1,0);
            }
            \foreach \pos in \verticalqubitsz {
                \draw[good_blue, line width=2pt] \pos -- ++(0,1);
            }
            
            \draw[fill=good_blue, color = good_blue] (0,0) circle (0.15);
        \end{scope}
    \end{tikzpicture}\;\;\;&     \;\;\;        
    \begin{tikzpicture}[scale=0.5]
        \def\horizontalqubits{(0,0),(2,1), (2,2)}
        \def\verticalqubits{(0,2),(1,2),(2,0)}

        \def\horizontalqubitsz{
        (\boxsize - 1 - 0,\boxsize - 1 - 2),
        (\boxsize - 1 - 1,\boxsize - 1 - 2),
        (\boxsize - 1 - 2,\boxsize - 1 - 0)
        }
        \def\verticalqubitsz{
        (\boxsize - 1 - 0,\boxsize - 1 - 0),
        (\boxsize - 1 - 2,\boxsize - 1 - 1),
        (\boxsize - 1 - 2,\boxsize - 1 - 2)}
        \def\boxsize{3}
        
        \begin{scope}[xshift=0cm]
            \draw[\latticecolor, \latticethickness] (3,0) -- (0,0) -- (0,3) ;
            \draw[\latticecolor, \latticethickness] (1,0) -- (1,3);
            \draw[\latticecolor, \latticethickness] (2,0) -- (2,3);
            \draw[\latticecolor, \latticethickness] (0,1) -- (3,1);
            \draw[\latticecolor, \latticethickness] (0,2) -- (3,2);
            \fill[color = good_red, opacity = .0] (0,0) -- (3,0) -- (3,3) -- (0,3) -- cycle;
            \foreach \pos in \horizontalqubits {
                \draw[good_red, line width=2pt] \pos -- ++(1,0);
            }
            \foreach \pos in \verticalqubits {
                \draw[good_red, line width=2pt] \pos -- ++(0,1);
            }
            
            \draw[fill=good_red, color = good_red] (0,0) circle (0.15);
        \end{scope}

        \def\horizontalqubits{(0,0),(0,1), (2,2)}
        \def\verticalqubits{(0,2),(1,0),(2,0)}
        \begin{scope}[xshift=4cm]
            \draw[\latticecolor, \latticethickness] (3,0) -- (0,0) -- (0,3) ;
            \draw[\latticecolor, \latticethickness] (1,0) -- (1,3);
            \draw[\latticecolor, \latticethickness] (2,0) -- (2,3);
            \draw[\latticecolor, \latticethickness] (0,1) -- (3,1);
            \draw[\latticecolor, \latticethickness] (0,2) -- (3,2);
            
            \foreach \pos in \horizontalqubitsz {
                \draw[good_blue, line width=2pt] \pos -- ++(1,0);
            }
            \foreach \pos in \verticalqubitsz {
                \draw[good_blue, line width=2pt] \pos -- ++(0,1);
            }
            
            \draw[fill=good_blue, color = good_blue] (0,0) circle (0.15);
        \end{scope}
    \end{tikzpicture} 
    
    \end{tabular}
    \caption{All pairs of weight 6 stabilizers in $3 \times 3$ boxes achieving the code parameters $[[288,8,12]]$. Eight pairs of $\op X$- (left) and $\op Z$-type (right) stabilizers are shown, and one can obtain the rest by swapping the role of $\op X$- and $\op Z$-type stabilizers.}
    \label{tab:w6_all}
\end{table}

\section{Stabilizers of weight 8 confined in $3 \times 3$ boxes yielding $[[288, 8, 14]]$ codes}\label{sec: appendix weight 8}

We exhaustively searched over all possible weight-8 stabilizers confined in a $3 \times 3$ tile. We find that there are more than 300 pairs of stabilizers yielding $[[288, 8,14]]$ codes. In \Cref{tab:w8_some}, we depicted some of them.
\begin{table}[h]
    \centering
    \begin{tabular}{c|c|c|c}
            \begin{tikzpicture}[scale=0.5]
        \def\horizontalqubits{(0,0),(0,1), (0,2), (2,0)}
        \def\verticalqubits{(0,0),(0,1),(1,1),(2,2)}

        \def\horizontalqubitsz{
            (\boxsize - 1 - 0,\boxsize - 1 - 0),
            (\boxsize - 1 - 0,\boxsize - 1 - 1),
            (\boxsize - 1 - 1,\boxsize - 1 - 1),
            (\boxsize - 1 - 2,\boxsize - 1 - 2)
        }
        \def\verticalqubitsz{
            (\boxsize - 1 - 0,\boxsize - 1 - 0),
            (\boxsize - 1 - 0,\boxsize - 1 - 1),
            (\boxsize - 1 - 0,\boxsize - 1 - 2),
            (\boxsize - 1 - 2,\boxsize - 1 - 0)
        }
        \def\boxsize{3}
        
        \begin{scope}[xshift=0cm]
            \draw[\latticecolor, \latticethickness] (3,0) -- (0,0) -- (0,3) ;
            \draw[\latticecolor, \latticethickness] (1,0) -- (1,3);
            \draw[\latticecolor, \latticethickness] (2,0) -- (2,3);
            \draw[\latticecolor, \latticethickness] (0,1) -- (3,1);
            \draw[\latticecolor, \latticethickness] (0,2) -- (3,2);
            \fill[color = good_red, opacity = .0] (0,0) -- (3,0) -- (3,3) -- (0,3) -- cycle;
            \foreach \pos in \horizontalqubits {
                \draw[good_red, line width=2pt] \pos -- ++(1,0);
            }
            \foreach \pos in \verticalqubits {
                \draw[good_red, line width=2pt] \pos -- ++(0,1);
            }
            
            \draw[fill=good_red, color = good_red] (0,0) circle (0.15);
        \end{scope}

        \def\horizontalqubits{(0,0),(0,1), (2,2)}
        \def\verticalqubits{(0,2),(1,0),(2,0)}
        \begin{scope}[xshift=4cm]
            \draw[\latticecolor, \latticethickness] (3,0) -- (0,0) -- (0,3) ;
            \draw[\latticecolor, \latticethickness] (1,0) -- (1,3);
            \draw[\latticecolor, \latticethickness] (2,0) -- (2,3);
            \draw[\latticecolor, \latticethickness] (0,1) -- (3,1);
            \draw[\latticecolor, \latticethickness] (0,2) -- (3,2);
            
            \foreach \pos in \horizontalqubitsz {
                \draw[good_blue, line width=2pt] \pos -- ++(1,0);
            }
            \foreach \pos in \verticalqubitsz {
                \draw[good_blue, line width=2pt] \pos -- ++(0,1);
            }
            
            \draw[fill=good_blue, color = good_blue] (0,0) circle (0.15);
        \end{scope}
        \addvmargin{1mm}
    \end{tikzpicture} \;\;\;&\;\;\;   
    \begin{tikzpicture}[scale=0.5]
        \def\horizontalqubits{(0,0),(0,1), (0,2), (2,0)}
        \def\verticalqubits{(0,0), (0,2), (1,1),(2,2)}

        \def\horizontalqubitsz{
        (\boxsize - 1 - 0,\boxsize - 1 - 0),
        (\boxsize - 1 - 0,\boxsize - 1 - 2),
        (\boxsize - 1 - 1,\boxsize - 1 - 1),
        (\boxsize - 1 - 2,\boxsize - 1 - 2)}
        \def\verticalqubitsz{
        (\boxsize - 1 - 0,\boxsize - 1 - 0),
        (\boxsize - 1 - 0,\boxsize - 1 - 1),
        (\boxsize - 1 - 0,\boxsize - 1 - 2),
        (\boxsize - 1 - 2,\boxsize - 1 - 0)}
        \def\boxsize{3}
        
        \begin{scope}[xshift=0cm]
            \draw[\latticecolor, \latticethickness] (3,0) -- (0,0) -- (0,3) ;
            \draw[\latticecolor, \latticethickness] (1,0) -- (1,3);
            \draw[\latticecolor, \latticethickness] (2,0) -- (2,3);
            \draw[\latticecolor, \latticethickness] (0,1) -- (3,1);
            \draw[\latticecolor, \latticethickness] (0,2) -- (3,2);
            \fill[color = good_red, opacity = .0] (0,0) -- (3,0) -- (3,3) -- (0,3) -- cycle;
            \foreach \pos in \horizontalqubits {
                \draw[good_red, line width=2pt] \pos -- ++(1,0);
            }
            \foreach \pos in \verticalqubits {
                \draw[good_red, line width=2pt] \pos -- ++(0,1);
            }
            
            \draw[fill=good_red, color = good_red] (0,0) circle (0.15);
        \end{scope}

        \def\horizontalqubits{(0,0),(0,1), (2,2)}
        \def\verticalqubits{(0,2),(1,0),(2,0)}
        \begin{scope}[xshift=4cm]
            \draw[\latticecolor, \latticethickness] (3,0) -- (0,0) -- (0,3) ;
            \draw[\latticecolor, \latticethickness] (1,0) -- (1,3);
            \draw[\latticecolor, \latticethickness] (2,0) -- (2,3);
            \draw[\latticecolor, \latticethickness] (0,1) -- (3,1);
            \draw[\latticecolor, \latticethickness] (0,2) -- (3,2);
            
            \foreach \pos in \horizontalqubitsz {
                \draw[good_blue, line width=2pt] \pos -- ++(1,0);
            }
            \foreach \pos in \verticalqubitsz {
                \draw[good_blue, line width=2pt] \pos -- ++(0,1);
            }
            
            \draw[fill=good_blue, color = good_blue] (0,0) circle (0.15);
        \end{scope}
    \end{tikzpicture} \;\;\;& \;\;\;             
    \begin{tikzpicture}[scale=0.5]
        \def\horizontalqubits{(0,0),(0,1), (0,2), (2,0)}
        \def\verticalqubits{(0,0),(2,1),(1,2),(2,2)}

        \def\horizontalqubitsz{
        (\boxsize - 1 - 0,\boxsize - 1 - 0),
        (\boxsize - 1 - 2,\boxsize - 1 - 1),
        (\boxsize - 1 - 1,\boxsize - 1 - 2),
        (\boxsize - 1 - 2,\boxsize - 1 - 2)
        }
        \def\verticalqubitsz{
        (\boxsize - 1 - 0,\boxsize - 1 - 0),
        (\boxsize - 1 - 0,\boxsize - 1 - 1),
        (\boxsize - 1 - 0,\boxsize - 1 - 2),
        (\boxsize - 1 - 2,\boxsize - 1 - 0)
        }
        \def\boxsize{3}
        
        \begin{scope}[xshift=0cm]
            \draw[\latticecolor, \latticethickness] (3,0) -- (0,0) -- (0,3) ;
            \draw[\latticecolor, \latticethickness] (1,0) -- (1,3);
            \draw[\latticecolor, \latticethickness] (2,0) -- (2,3);
            \draw[\latticecolor, \latticethickness] (0,1) -- (3,1);
            \draw[\latticecolor, \latticethickness] (0,2) -- (3,2);
            \fill[color = good_red, opacity = .0] (0,0) -- (3,0) -- (3,3) -- (0,3) -- cycle;
            \foreach \pos in \horizontalqubits {
                \draw[good_red, line width=2pt] \pos -- ++(1,0);
            }
            \foreach \pos in \verticalqubits {
                \draw[good_red, line width=2pt] \pos -- ++(0,1);
            }
            
            \draw[fill=good_red, color = good_red] (0,0) circle (0.15);
        \end{scope}

        \def\horizontalqubits{(0,0),(0,1), (2,2)}
        \def\verticalqubits{(0,2),(1,0),(2,0)}
        \begin{scope}[xshift=4cm]
            \draw[\latticecolor, \latticethickness] (3,0) -- (0,0) -- (0,3) ;
            \draw[\latticecolor, \latticethickness] (1,0) -- (1,3);
            \draw[\latticecolor, \latticethickness] (2,0) -- (2,3);
            \draw[\latticecolor, \latticethickness] (0,1) -- (3,1);
            \draw[\latticecolor, \latticethickness] (0,2) -- (3,2);
            
            \foreach \pos in \horizontalqubitsz {
                \draw[good_blue, line width=2pt] \pos -- ++(1,0);
            }
            \foreach \pos in \verticalqubitsz {
                \draw[good_blue, line width=2pt] \pos -- ++(0,1);
            }
            
            \draw[fill=good_blue, color = good_blue] (0,0) circle (0.15);
        \end{scope}
    \end{tikzpicture}\;\;\;&   \;\;\;          
    \begin{tikzpicture}[scale=0.5]
        \def\horizontalqubits{(0,0),(0,1), (0,2), (2,0)}
        \def\verticalqubits{(0,1),(1,0),(1,1),(2,2)}

        \def\horizontalqubitsz{
        (\boxsize - 1 - 0,\boxsize - 1 - 1),
        (\boxsize - 1 - 1,\boxsize - 1 - 0),
        (\boxsize - 1 - 1,\boxsize - 1 - 1),
        (\boxsize - 1 - 2,\boxsize - 1 - 2)}
        \def\verticalqubitsz{
        (\boxsize - 1 - 0,\boxsize - 1 - 0),
        (\boxsize - 1 - 0,\boxsize - 1 - 1),
        (\boxsize - 1 - 0,\boxsize - 1 - 2),
        (\boxsize - 1 - 2,\boxsize - 1 - 0)}
        \def\boxsize{3}
        
        \begin{scope}[xshift=0cm]
            \draw[\latticecolor, \latticethickness] (3,0) -- (0,0) -- (0,3) ;
            \draw[\latticecolor, \latticethickness] (1,0) -- (1,3);
            \draw[\latticecolor, \latticethickness] (2,0) -- (2,3);
            \draw[\latticecolor, \latticethickness] (0,1) -- (3,1);
            \draw[\latticecolor, \latticethickness] (0,2) -- (3,2);
            \fill[color = good_red, opacity = .0] (0,0) -- (3,0) -- (3,3) -- (0,3) -- cycle;
            \foreach \pos in \horizontalqubits {
                \draw[good_red, line width=2pt] \pos -- ++(1,0);
            }
            \foreach \pos in \verticalqubits {
                \draw[good_red, line width=2pt] \pos -- ++(0,1);
            }
            
            \draw[fill=good_red, color = good_red] (0,0) circle (0.15);
        \end{scope}

        \def\horizontalqubits{(0,0),(0,1), (2,2)}
        \def\verticalqubits{(0,2),(1,0),(2,0)}
        \begin{scope}[xshift=4cm]
            \draw[\latticecolor, \latticethickness] (3,0) -- (0,0) -- (0,3) ;
            \draw[\latticecolor, \latticethickness] (1,0) -- (1,3);
            \draw[\latticecolor, \latticethickness] (2,0) -- (2,3);
            \draw[\latticecolor, \latticethickness] (0,1) -- (3,1);
            \draw[\latticecolor, \latticethickness] (0,2) -- (3,2);
            
            \foreach \pos in \horizontalqubitsz {
                \draw[good_blue, line width=2pt] \pos -- ++(1,0);
            }
            \foreach \pos in \verticalqubitsz {
                \draw[good_blue, line width=2pt] \pos -- ++(0,1);
            }
            
            \draw[fill=good_blue, color = good_blue] (0,0) circle (0.15);
        \end{scope}
    \end{tikzpicture}     
    \end{tabular}
    \caption{Four selected pairs of weight 8 stabilizers in $3 \times 3$ boxes achieving the code parameters $[[288,8,14]]$.}
    \label{tab:w8_some}
\end{table}
\end{widetext}

\end{document}